\documentclass[12pt]{article}
\RequirePackage[top=1in, left=1in, right=1in]{geometry}

\usepackage{blindtext}
\usepackage[english]{babel}
\usepackage{natbib}
\usepackage{amsmath,amssymb,amstext,amsthm,graphics, layout, lscape, longtable}
\usepackage[normalem]{ulem}
\usepackage{color}

\usepackage{url}
\usepackage{caption}
\usepackage{amssymb,amsmath,graphicx,multicol,epsfig,pictexwd,spverbatim}
\usepackage{color, colortbl}
\usepackage{multirow,multicol,dcolumn,booktabs}
\definecolor{mygray}{gray}{0.8}
\usepackage{lscape}
\usepackage[utf8]{inputenc}
\usepackage{hyperref}

\usepackage{tablefootnote}
\usepackage{tabu}
\usepackage{lscape}
\usepackage{array}
\usepackage{subfig}
\usepackage{caption}
\usepackage{bm}
\usepackage{xcolor,colortbl}
\usepackage{multirow}
\usepackage{fourier} 
\usepackage{array}
\usepackage{makecell}
\usepackage{longtable}
\usepackage{authblk}

\usepackage{array}
\usepackage{xcolor}

\makeatletter

\newcommand*{\@rowstyle}{}

\newcommand*{\rowstyle}[1]{
  \gdef\@rowstyle{#1}%
  \@rowstyle\ignorespaces%
}

\newcolumntype{=}{
  >{\gdef\@rowstyle{}}%
}

\newcolumntype{+}{
  >{\@rowstyle}%
}

\makeatother
\newcommand{\bi}{\begin{itemize}}
\newcommand{\ei}{\end{itemize}}

\newcolumntype{C}{@{\extracolsep{.4em}}c@{\extracolsep{0pt}}}%

\title{Validation of a Bayesian Learning Model to Predict the Risk for Cannabis Use Disorder}
\author[a]{Thanthirige Lakshika M. Ruberu}
\author[a]{Rajapaksha Mudalige Dhanushka S. Rajapaksha}
\author[b]{Mary M. Heitzeg}
\author[b]{Ryan Klaus}
\author[c]{Joseph M. Boden}
\author[a$^{*}$]{Swati Biswas}
\author[a$^{*}$]{Pankaj Choudhary}

\affil[a]{Department of Mathematical Sciences, University of Texas at Dallas, Richardson, TX 75080, United States}
\affil[b]{Department of Psychiatry, University of Michigan, Ann Arbor, MI 48109, United States}
\affil[c]{Department of Psychological Medicine, University of Otago, Christchurch 8011, New Zealand}
\date{}
\begin{document}
\maketitle
*Correspondence to Swati Biswas, 800 W Campbell Rd, FO 35, Richardson, TX 75025. Email: swati.biswas@utdallas.edu or Pankaj Choudhary 800 W Campbell Rd, FO 35, Richardson, TX 75025. Email: pankaj@utdallas.edu

\newpage
\begin{abstract}
\noindent
\textbf{Background:} Cannabis use disorder (CUD) is a growing public health problem. Early identification of adolescents and young adults at risk of developing CUD in the future may help stem this trend. A logistic regression model fitted using a Bayesian learning approach was developed recently to predict the risk of future CUD based on seven risk factors in adolescence and youth. A nationally representative longitudinal dataset, Add Health was used to train the model (henceforth referred as Add Health model).
\noindent$\\$
\textbf{Methods:} We validated the Add Health model on two cohorts, namely, Michigan Longitudinal Study (MLS) and Christchurch Health and Development Study (CHDS) using longitudinal data from participants until they were approximately 30 years old (to be consistent with the training data from Add Health). If a participant was diagnosed with CUD at any age during this period, they were considered a case. We calculated the area under the curve (AUC) and the ratio of expected and observed number of cases (E/O). We also explored re-calibrating the model to account for differences in population prevalence.
\noindent$\\$
\textbf{Results:} The cohort sizes used for validation were 424 (53 cases) for MLS and 637 (105 cases) for CHDS. AUCs for the two cohorts were 0.66 (MLS) and 0.73 (CHDS) and the corresponding E/O ratios (after recalibration) were 0.995 and 0.999.
\noindent$\\$
\textbf{Conclusion:} The external validation of the Add Health model on two different cohorts lends confidence to the model's ability to identify adolescent or young adult cannabis users at high risk of developing CUD in later life.
\end{abstract}

Keywords: Add Health, Michigan Longitudinal Study, Christchurch Health and Development Study, Model discrimination, Model calibration

\newpage
\section{Introduction}
Substance use disorders (SUDs) are one of the most serious and costly public health issues in the US. In 2020, 40.3 million (14.5\%) people aged 12 or older  had an SUD in the past year \citep{abuse2020key}. National Drug Control Budget obtained \$34.6 billion for drug control including treatment, which amounted to 45\% of their total budget \citep{ncdas_2022}.

Cannabis is the most commonly used illicit drug in the US with 49.6 million past year users among people aged 12 or older \citep{abuse2020key} and its usage may lead to cannabis use disorder (CUD), the most prevalent illicit drug use disorder. The 2020 National Survey on Drug Use and Health reported that among 18.4 million people (aged 12 or older) diagnosed with illicit drug disorder in the past year, 14.2 million were diagnosed with CUD \citep{abuse2020key}. Moreover, 30\%  of those who use cannabis have been found to report at least one or more symptoms of CUD, irrespective of whether they met criteria for the disorder \citep{ncdas_2022}.

The development of CUD is thought to be the result of the interplay of a large variety of individual, family and social factors that work to increase the CUD risk for an individual. For example, data from the Christchurch Health and Development Study (CHDS) suggests that family influences such as parental illicit drug use, exposure to abuse in childhood, a personality profile high in novelty-seeking, being male, and peer influence are key factors in the development of CUD \citep{boden2006illicit}. In addition, other research suggests that cannabis initiation at an earlier age, and progression to more regular use at earlier ages increases the risk of developing CUD at a later point in life \citep{silins2014young}. There is also some evidence suggesting that the presence of a co-occurring mood disorder (not including major depression) can increase the risk of transitioning from cannabis use to CUD \citep{wittchen2007cannabis}. Finally, there has been speculation in the literature that changes in legal frameworks toward a more liberalized approach to cannabis may contribute to an increased risk of CUD in some population groups, but data thus far have been lacking \citep{taylor2019determining}.

Rates of cannabis use and CUD are of particular concern among young people (adolescents and young adults), who may be particularly vulnerable to the adverse effects of cannabis use \citep{blest2020adolescent}. Evidence suggests that in recent decades, there has been a gradual increase in past year cannabis use over time in the US population aged 12 or older \citep{abuse2020key}. In 2019, the National Center for Drug Abuse Statistics  \citep{ncdas_2022} reported that 43$\%$ of college students indicated having used cannabis in the past 12 months, and that among college students using illicit drugs, the vast majority (93$\%$) report using cannabis. Furthermore, the NCDAS reported that of those who used cannabis before age 18, one in six went on to develop CUD. The uptake and persistence of cannabis use among young people may also be impacted by the increasing popularity of cannabis vaping. For example, 62.8$\%$ of 12$^{\text{th}}$ graders who used cannabis in the past year reported doing so via vaping \citep{ncdas_2022_youth}. These findings suggest that an imperative for research on the uptake and persistence of cannabis use among young people should be to develop methods to identify those adolescent and young adult substance users who are at risk of developing CUD (or a substance use disorder more generally).

To aid clinicians/counselors in this goal, \citet{rajapaksha2022bayesian} recently developed a quantitative model that predicts the risk of developing CUD in adulthood. It is a multiple logistic regression model fitted using a novel Bayesian learning approach and based on data from National Longitudinal Study of Adolescent to Adult Health (Add Health, $n = 8712$), a nationally representative longitudinal study (Harris et al., 2009)\nocite{harriss2009}. The model consists of seven predictors: biological sex; scores on personality traits of neuroticism, openness, and conscientiousness; and measures of adverse childhood experiences, delinquency, and peer cannabis use. In particular, higher likelihood of CUD is associated with male sex, greater adverse childhood experiences and involvement in delinquent activities, higher neuroticism and openness, lower conscientiousness, and larger proportion of friends using cannabis. The model was also validated on an independent sample from Add Health. To the best of our knowledge, there is no other CUD risk prediction model that is based on nationally representative longitudinal data, which also has been externally validated.            

In the current work, we  validated the model developed in \citet{rajapaksha2022bayesian}  on two independent and external data sets, one from the Michigan Longitudinal Study (MLS)  \citep{zucker1996other, zucker2002clinical} and the other from the Christchurch Health and Development Study (CHDS) \citep{poulton2020patterns}. We chose these two datasets because they are longitudinal studies with data available on the seven risk factors needed by the model of  \citet{rajapaksha2022bayesian}, referred to as Add Health model henceforth.

\section{Methods}
\subsection{Validation Datasets}
\label{paricipants}
MLS monitored indicators of child, parent, family, peer group, school, and other environmental influences on risk and protective behaviours over the life course of its participants since it began in 1989 \citep{zucker1996other, zucker2002clinical}. Participants were from families ascertained through two population-based ecologically comparable samples, with and without father's alcoholism. Father's drunk driving conviction records were used to ascertain families with alcoholism (high-risk families) while the families without father's alcoholism were recruited through door-to-door communication canvassing the same neighbourhood as the high-risk families. For initial recruitment, the study required father to be living with a 3-5 year old male child and biological mother of the child. Approximately two years after the initial recruitment, siblings within 8 years of age of the initially recruited male child were also recruited. MLS collected data on participants from ages 3-5 years through 39-41 ages in two types of waves: 3-year waves up to age 21 and 6-year waves afterwards. Additionally, two separate short annual assessments were conducted between ages 11-17 and 18-26. The validation sample included 424 cannabis users.

CHDS collected data on health, education, and life progress of a group of 1,265 children born in Christchurch, New Zealand during 1977 \citep{poulton2020patterns}. The cohort represented 97$\%$ of births in the Christchurch urban region of New Zealand between April and
August 1977 and has been studied at birth, 4 months, 1 year, annually up to age 16, and at ages 18, 21, 25, 30, 35, and 40 years. The final sample used to validate the Add Health model consisted of 637 cannabis users.

\subsection{Measures Used in Model Validation}
\label{details}
The seven risk factors needed by the Add Health model are described in Table \ref{VAr comparison}. Of these, delinquency and peer cannabis use are longitudinal variables. The model internally uses summarized versions of delinquency score in terms of average over the waves in which it was measured and peer cannabis use in terms of estimated random effects (as described in \citet{rajapaksha2022bayesian}). Those summaries are used in the model in a similar way as the other (cross-sectional) variables. All variables (except biological sex) need to be in the scale of 0 to 1. \citet{rajapaksha2022bayesian} used this scale to ensure portability of the model across datasets as different datasets may measure the variables in different scales. The response variable in the Add Health model is a binary variable of CUD diagnosis based on Diagnostic and Statistical Manual for Mental Disorders (DSM) IV  \citep{ap2000diagnostic} criteria. We tried to match the Add Health response variable with corresponding measures from MLS and CHDS to the best of our ability. The specific details on matching the response variable can be found below. The details on the matching of seven risk factors are summarized in Table \ref{VAr comparison} and further elaborated in Supplementary Materials section. As in \citet{rajapaksha2022bayesian}, to ensure that all predictors were measured before the outcome of CUD, we took into account the age at data collection for each risk factor, except for the three personality variables. This exception was made because CHDS only collected data on conscientiousness scale at age 40. However, these personality traits are considered to be relatively stable over time \citep{caspi2005personality,damian2019sixteen}.

\subsubsection{Cannabis Use Disorder}
In Add Health cohort, CUD was measured retrospectively when the participants were between ages 24-32. That is, CUD could have occurred before that age.
MLS assessed DSM IV criteria using Diagnostic Interview Schedule (DIS) \citep{robins1995diagnostic} at every 3-year assessment. Any participant diagnosed with CUD measured between waves 4 (ages 12 -- 16) to 9 (ages 26 -- 32) was considered a case. By wave 9, the participants were around age 30 and it matched with the above-mentioned time frame in Add Health (there were no CUD cases before wave 4).

CHDS had past year DSM IV dependence measured at each year from 15 to 30 years using a number of questionnaires including the alcohol use questionnaires developed by Casswell and colleagues \citep{casswell1991longitudinal,mccrae1992revised} and the Composite Diagnostic Interview Schedule \citep{CIDI}. A participant was considered a case if they were diagnosed with CUD at any age between 15 to 30.

\subsection{Statistical Methods}
\subsubsection{Details about the Add Health Model}
The Add Health model estimates for a cannabis user the probability of developing CUD in the adulthood based on adolescent personal risk factors. Note that being on probability scale, the numeric value gives an indication of the severity of risk of developing CUD. Let $y_i$ be the CUD status of the $i^{th}$ participant, with $y_i=1,0$ for a case and a control, respectively, $i=1,\dots ,n $. It is assumed that
$y_i |p_i \sim \text{independent Bernoulli} (p_i)$ where $p_i$ is estimated using  a logistic regression model. The model includes both cross-sectional and longitudinal predictors. Two possibilities for including longitudinal predictors were explored. First approach was simply computing the average over the waves in which the predictor was measured and second approach was computing participant-specific random intercept associated with the predictor. The final model used the average across the waves for delinquency and random intercept for peer cannabis use. The regression coefficients were estimated by a Bayesian learning approach after specifying suitable priors. Technical details of the model can be found in \citet{rajapaksha2022bayesian}.
\subsubsection{Validation Methods}
\label{stat}
The predictive performance of the Add Health model in the two external validation datasets was assessed by computing discrimination and calibration measures. We assessed model discrimination using the Receiver Operating Curve (ROC) and its corresponding Area Under the Curve (AUC). Estimate of AUC and its 95$\%$ confidence interval were computed using pROC package in R \citep{proc,Rstudio}.  The calibration of the model was measured by comparing the expected (E) number of CUD cases (as predicted by the model) with the observed (O) number and via a calibration plot, which graphically represents the agreement between the observed and predicted probabilities/counts \citep{janssen2008updating}. Technical details about the calibration plot can be found in Supplementary Materials section. 

Due to differences in the prevalence of CUD between Add Health and the two validation data sets, we also considered re-calibrating the model \citep{moons2012risk}. In particular, the simplest way is to update the intercept of the Add Health model to reflect correct prevalence (the coefficients of the predictors remain unchanged). Note that updating the intercept of the Add Health model does not change the AUC because the relative ranking of the predicted probabilities stays the same. Details about model recalibration can be found in Supplementary Materials section.

\section{Results}

\subsection{Sample Characteristics}
\begin{figure}[htp]
\subfloat[ACE scale]{\includegraphics[width = 2.1in]{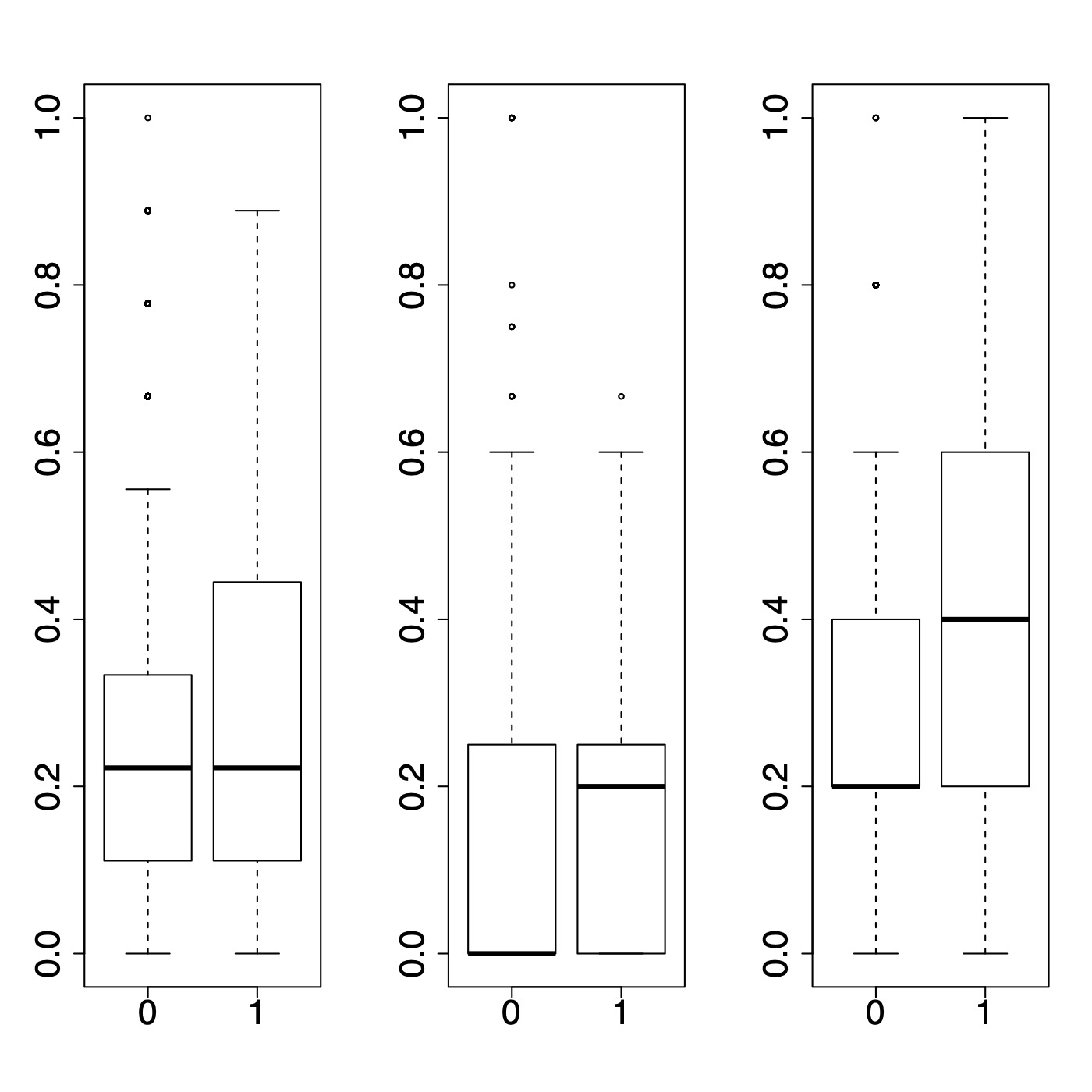}} \hspace{1.7em}
\subfloat[Neuroticism scale ]{\includegraphics[width = 2.1in]{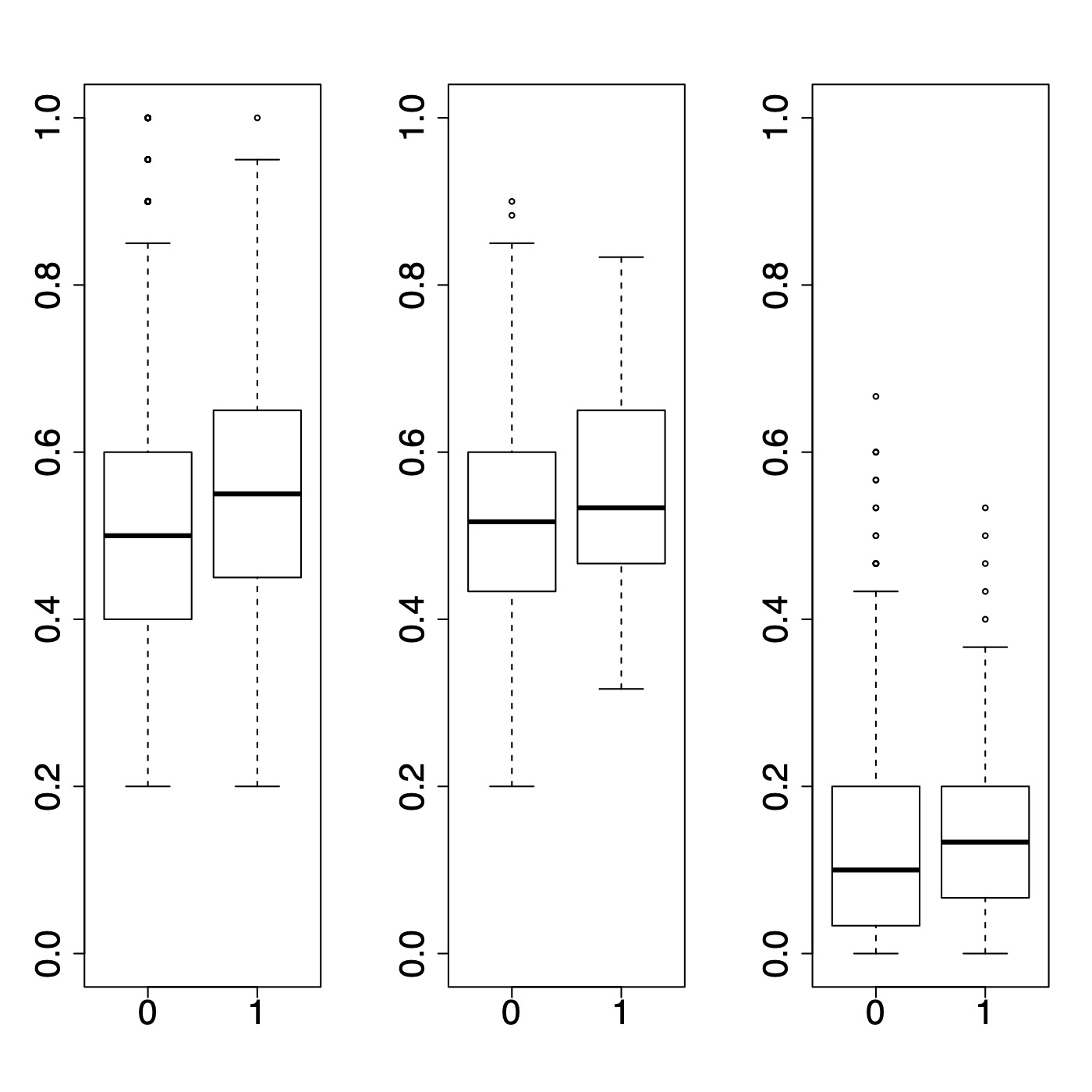}}\hspace{1.7em}
\subfloat[Conscientiousness scale]{\includegraphics[width = 2.1in]{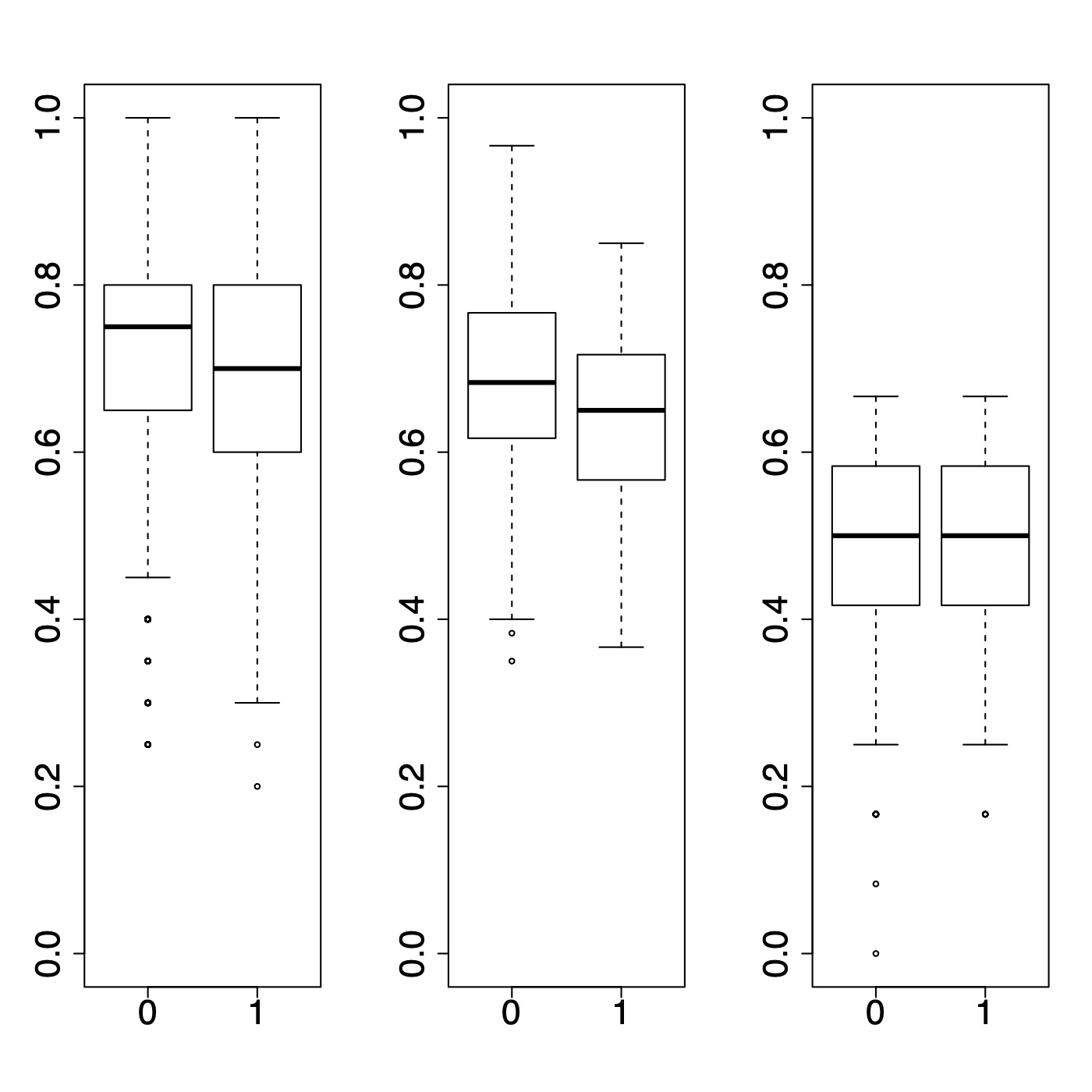}}\\
\subfloat[Openness scale]{\includegraphics[width = 2.1in]{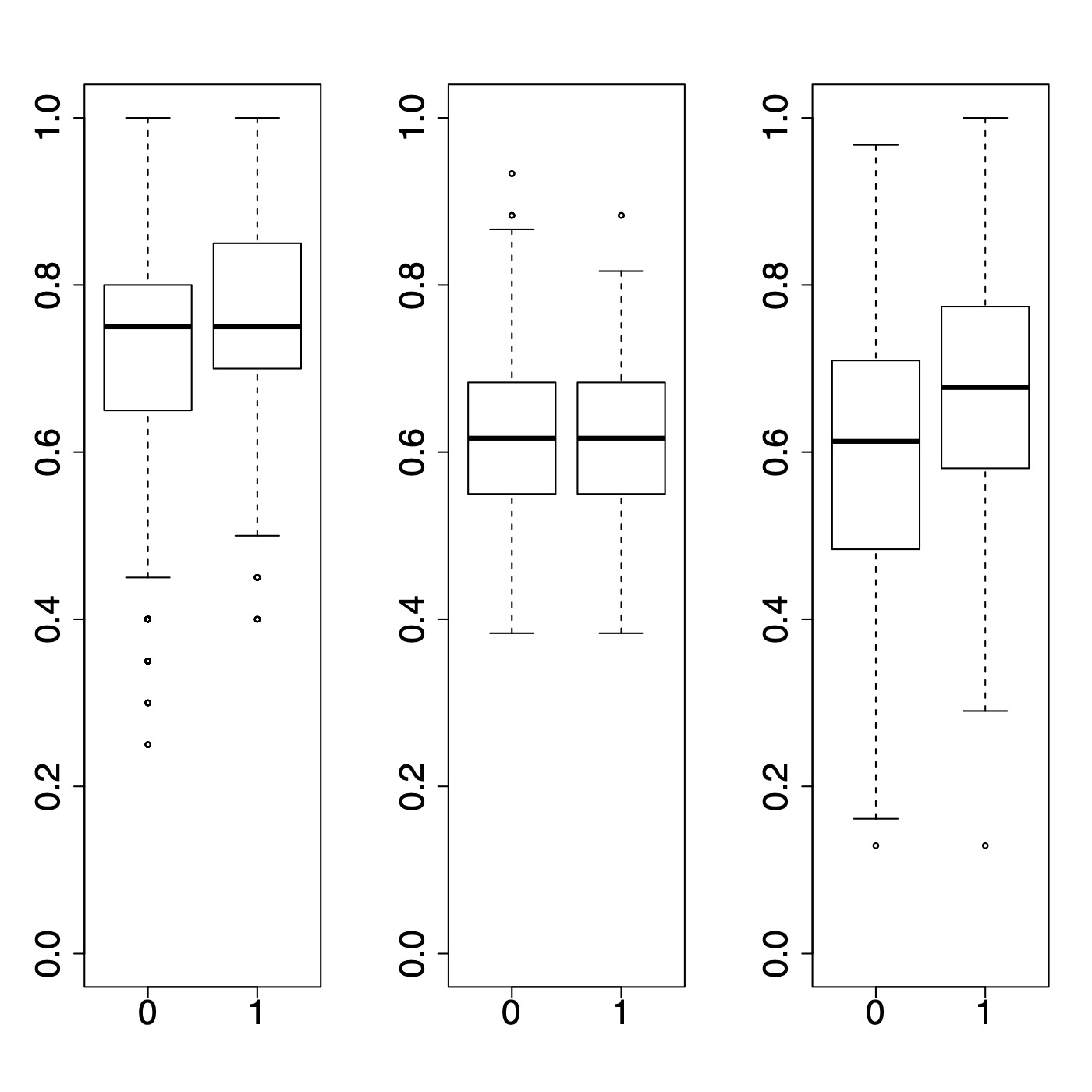}} \hspace{1.7em}
\subfloat[Delinquency]{\includegraphics[width = 2.1in]{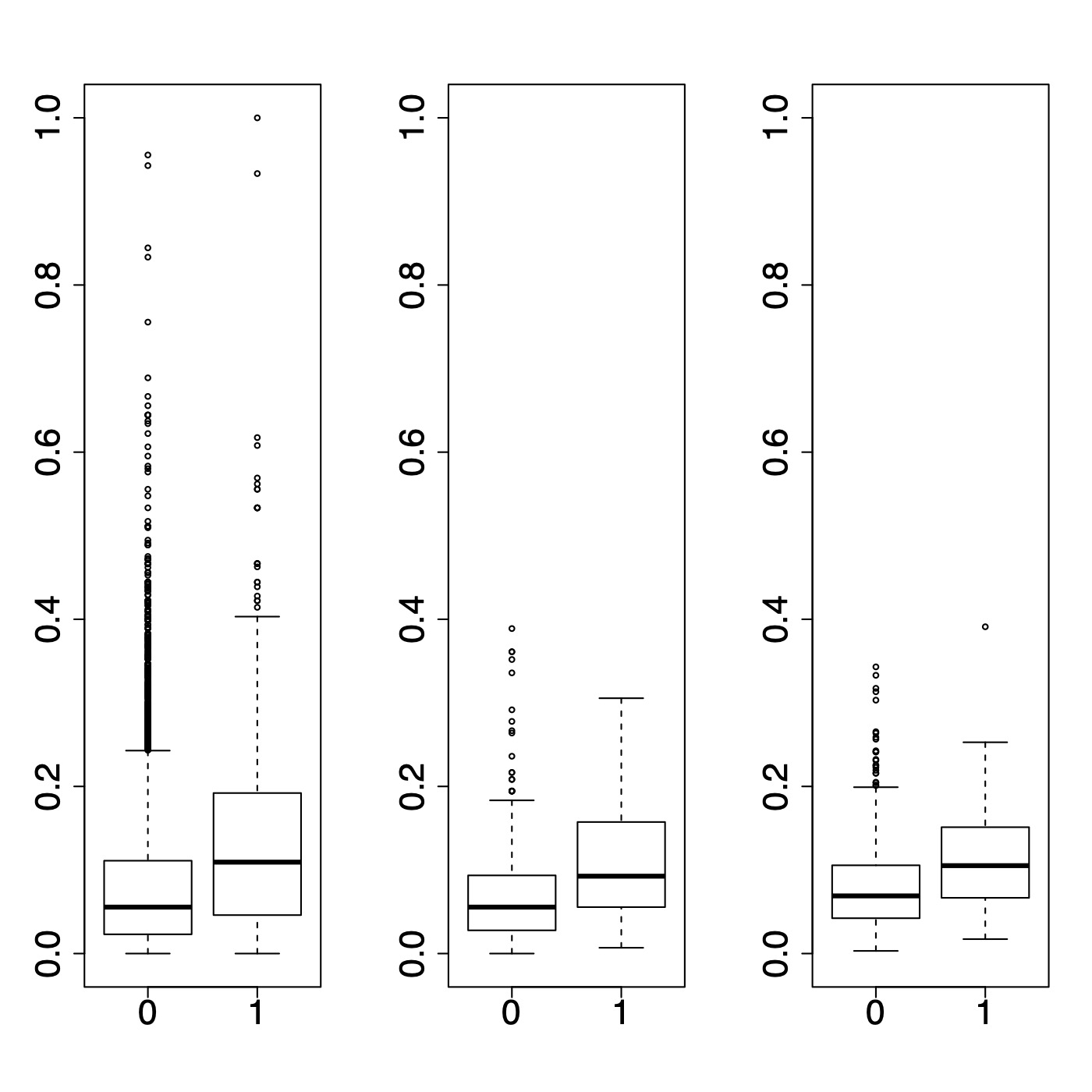}}\hspace{1.7em}
\subfloat[Peer cannabis use]{\includegraphics[width = 2.1in]{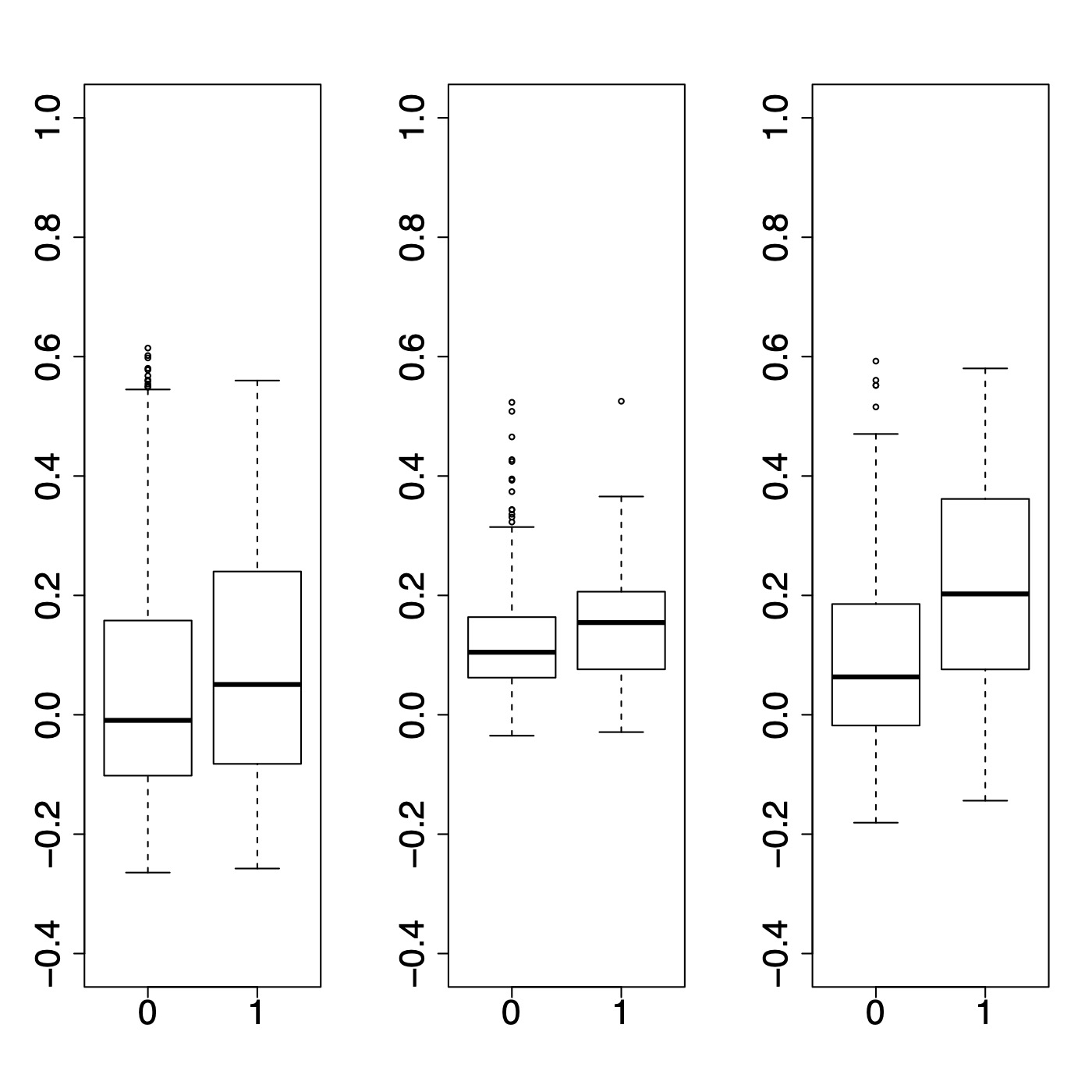}}\\
\caption{Boxplots of continuous predictors in Add Health (left), MLS (center), and CHDS (right). All predictor values except peer cannabis use are in standardized (0-1) scale; for peer cannabis use, distribution of random effects is shown; 0 and 1 on the x-axes indicate CUD cases and controls, respectively.}
\label{compare}
\end{figure}

To get an insight into the distribution of variables in MLS and CHDS and how they compare to each other and with those of Add Health, we show side-by-side box plots Figure ~\ref{compare}.  

\subsubsection{MLS Data}
In the final dataset ($n = 424$), the lifetime prevalence of CUD was 12.50\%. As seen in Table ~\ref{summary_cross_MLS}, cases had a higher proportion of males compared to controls (participants not diagnosed with CUD) (77.36\% vs 69.27\%) , experienced more ACEs (0.20 vs 0.18), and reported higher neuroticism (0.55 vs 0.51) but lower conscientiousness (0.64 vs 0.69). Openness scores were similar between cases and controls. Following Section ~\ref{details}, we only provide the mean and standard deviation of the average delinquency score and random effect estimates for peer cannabis use measures over all waves in Table \ref{summary_cross_MLS}. For both of these predictors, the average scores were higher for cases compared to controls.

\subsubsection{CHDS Data}
Table ~\ref{summary_cross_CHDS} summarizes the sample characteristics of CHDS data ($n = 637$). The lifetime prevalence of CUD among cannabis users was 16.48\%, which is higher compared to both Add Health ($7.5\%$) and MLS data. The male-to-female ratio was almost 1  unlike in MLS data wherein the majority of the sample were males. However, in CHDS there were more males among cases than in controls (70.48\% vs 46.05\%). The mean differences between cases and controls for all the predictors were in the same direction as in Add Health and MLS data sets except for openness scale. Similar to Add Health and MLS data, cases experienced more ACEs (0.41 vs 0.34), reported higher neuroticism (0.16 vs 0.14), lower conscientiousness (0.46 vs 0.49), higher average of delinquency values (0.12 vs 0.08), and larger random effects scores for peer cannabis use (0.22 vs 0.10). For openness scale, cases reported higher values compared to controls (0.68 vs 0.60)   unlike in MLS and Add Health datasets, in which, the cases and controls had similar means.  

\subsection{Validation Results}
The AUC values were 0.66 (0.58-0.74) and 0.73 (0.68-0.78) for MLS and CHDS datasets, respectively. The corresponding E/O ratios were 0.512 and 0.30. That is, the model under-predicted the number of cases in both datasets. The estimated calibration intercepts were 0.76 for MLS and 1.40 for CHDS, confirming that the model indeed under-predicts the number of cases (recall the ideal value is 0). 

\begin{figure}[htp]
\subfloat[Before Intercept Update]{\includegraphics[width = 3.5in]{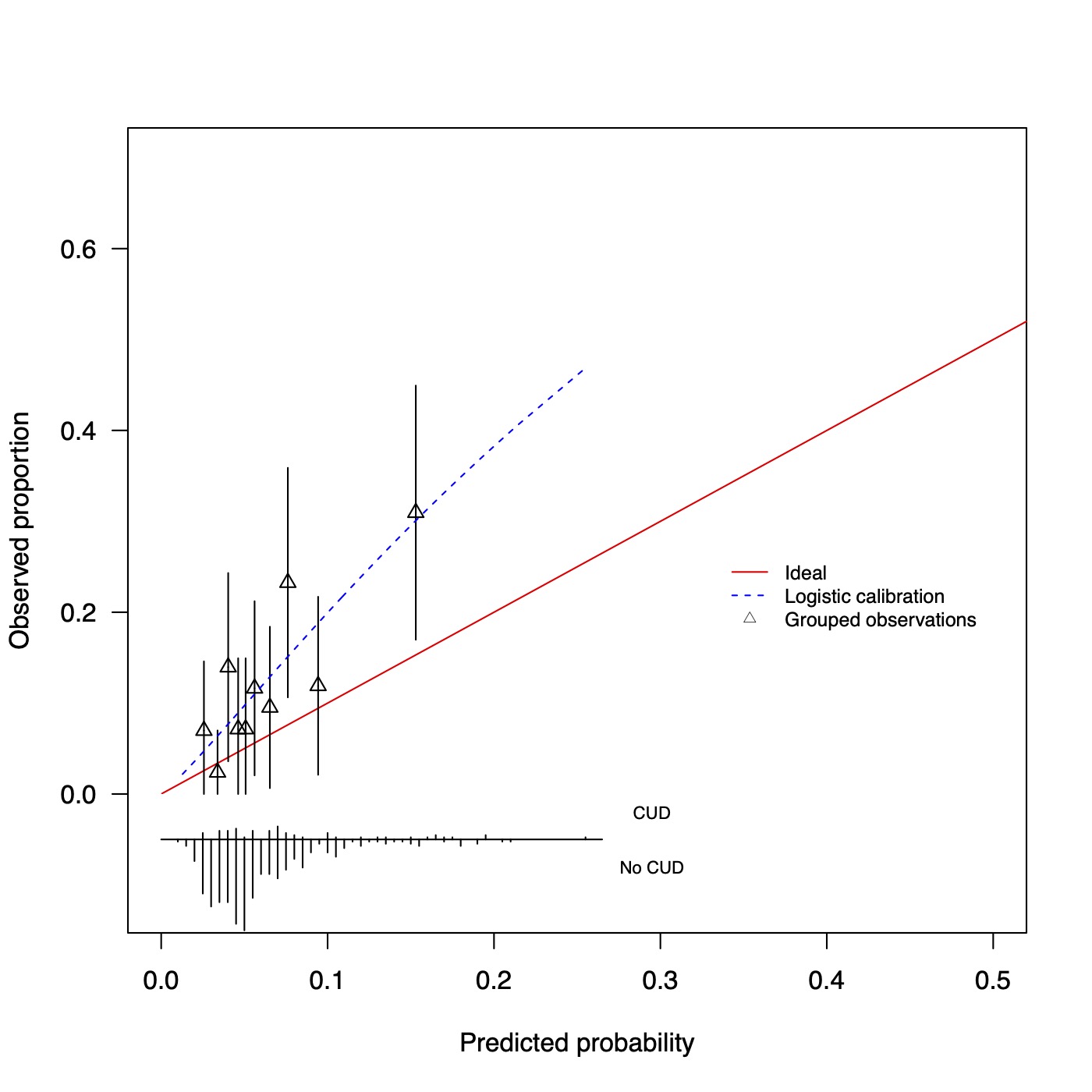}} \hspace{1em}
\subfloat[After Intercept Update]{\includegraphics[width = 3.5in]{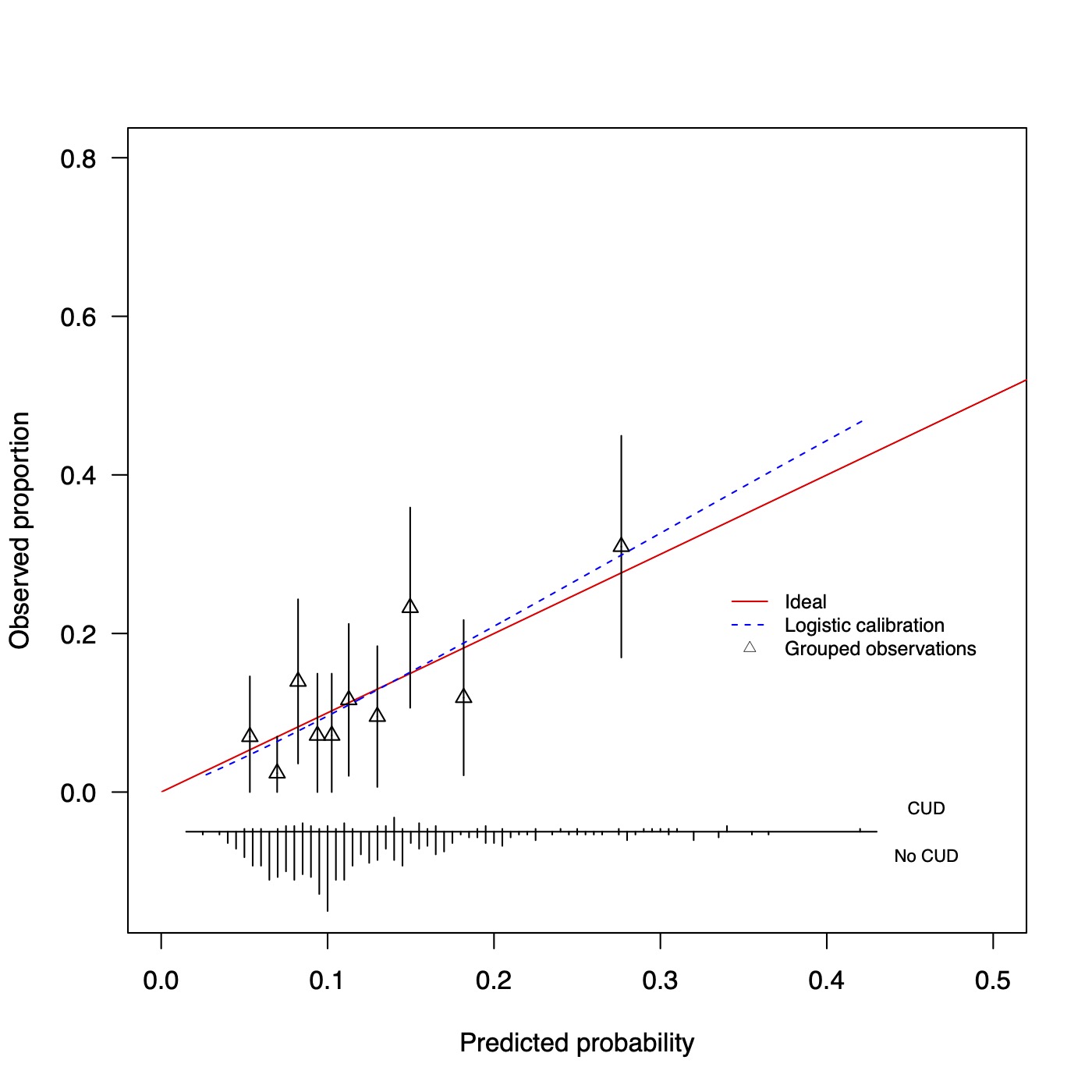}}

\caption{Comparison of calibration curves for MLS data. Perfect calibration
is represented by the straight line (red) through the origin with slope equal to 1. Triangles are based on deciles
of subjects with similar predicted probabilities for developing CUD. The frequency distribution of predicted probabilities is shown right above
the x-axis (as vertical bars) separately for cases and controls. }
\label{MLSfigure}
\end{figure}

\begin{figure}[htp]
\subfloat[Before Intercept Update]{\includegraphics[width = 3.5in]{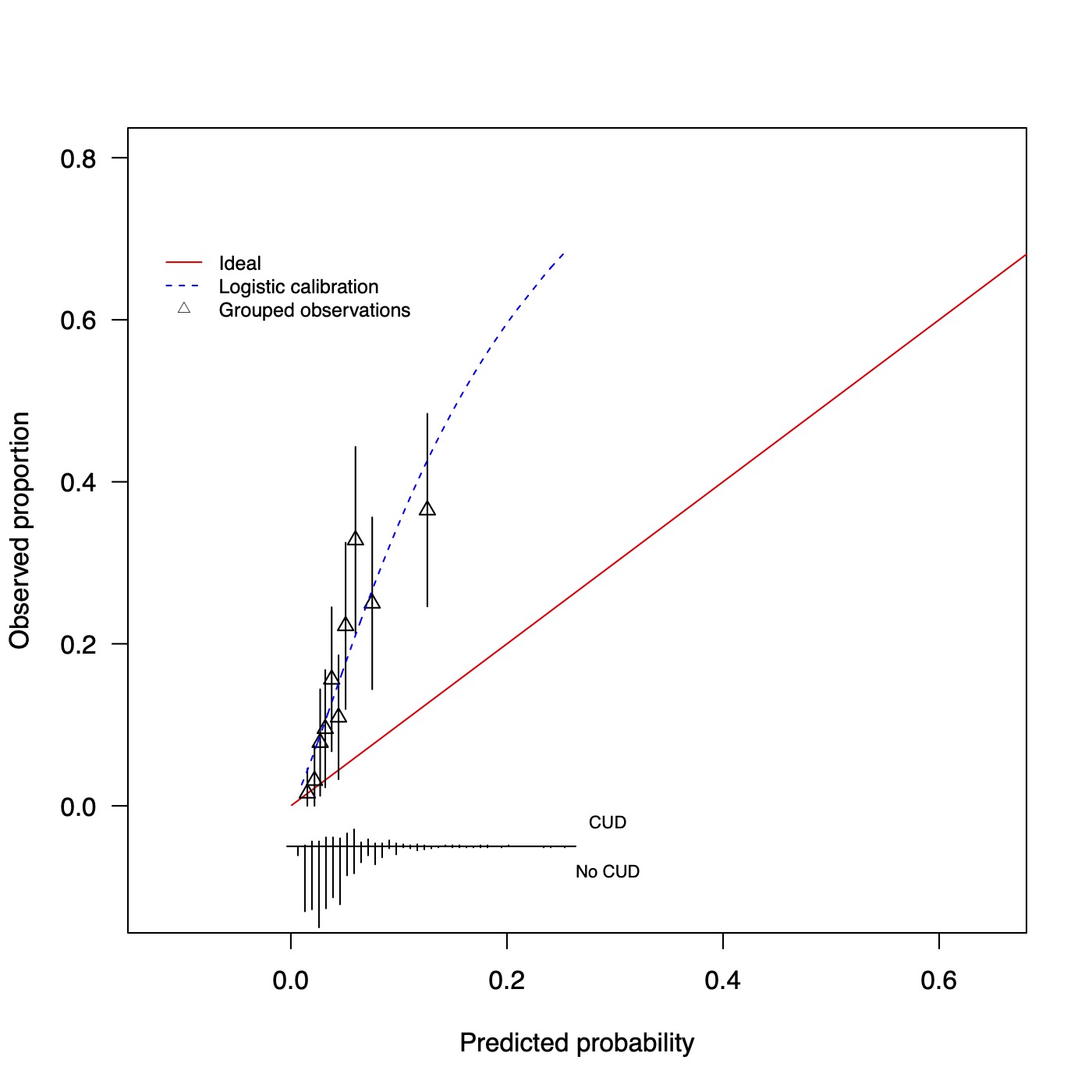}} \hspace{1em}
\subfloat[After Intercept Update]{\includegraphics[width = 3.5in]{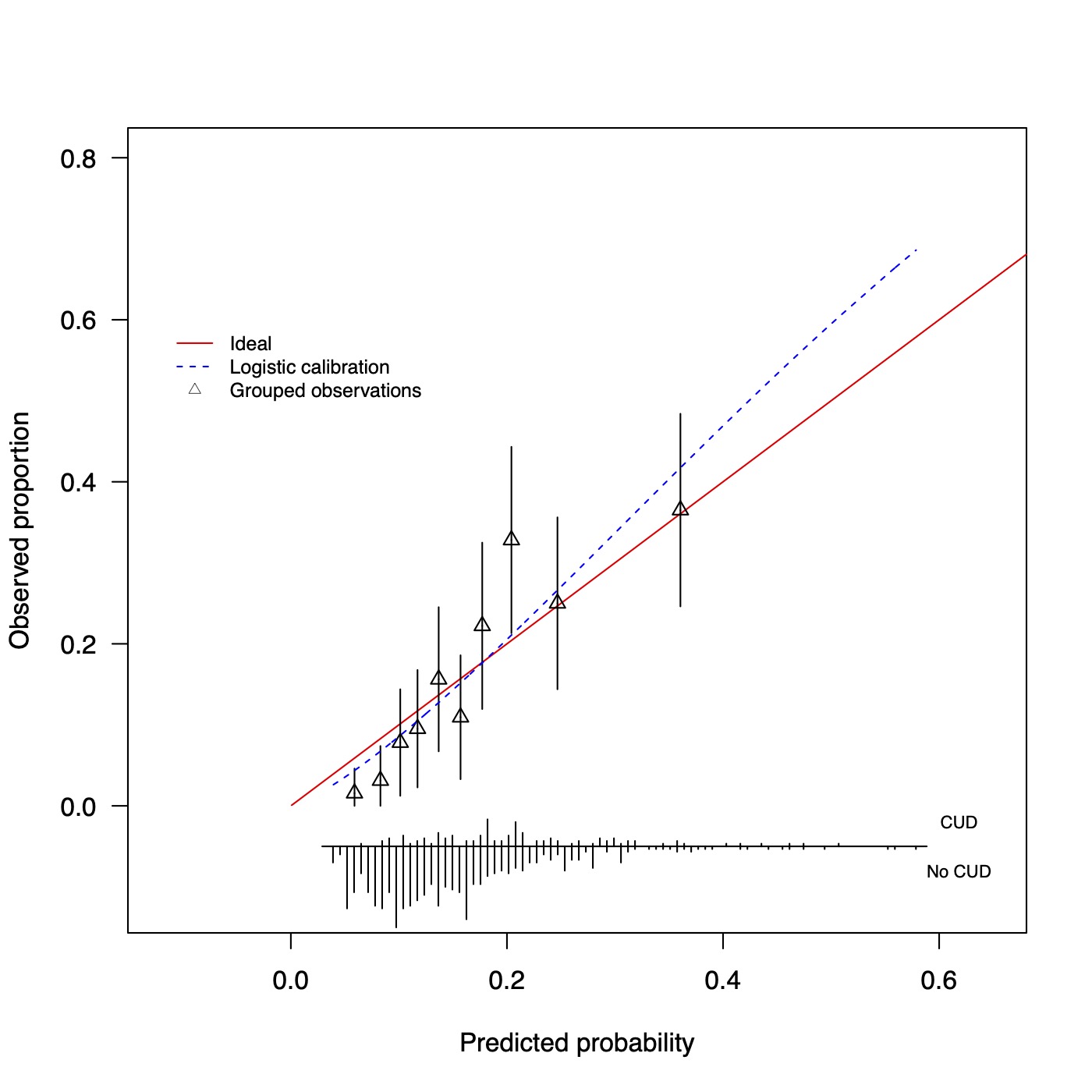}}

\caption{Comparison of calibration curves for CHDS data. Perfect calibration
is represented by the straight line (red) through the origin with slope equal to 1. Triangles are based on deciles
of subjects with similar predicted probabilities for developing CUD. The frequency distribution of predicted probabilities is shown right above
the x-axis (as vertical bars) separately for cases and controls. }
\label{CHDSfigure}
\end{figure}

As the baseline prevalences of these two datasets were substantially different from that in Add Health (this point will be elaborated in discussion), we attempted to re-calibrate the model (Section \ref{stat}). After adjusting the intercept of the Add Health model by adding the estimated calibration intercepts, the calibration of the model improved significantly. The updated E/O ratio was 0.999 for MLS data and 0.995 for CHDS data. Figures \ref{MLSfigure} and \ref{CHDSfigure} show the  calibration plots before and after updating the intercept for MLS and CHDS data, respectively. Tables \ref{quantiles_MLS} and \ref{quantiles_CHDS}  show the updated E/O values for the five risk quantile groups, two levels of biological sex, and the groups obtained by using median as a cutoff for each risk factor for both the data sets. Note that AUC values remained unchanged with the model recalibration as expected.


\section{Discussion}
 Given the growing prevalence of CUD in US \citep{abuse2020key}, it is imperative to identify adolescents and youth who are at risk of developing CUD in adulthood so that they can be helped early to mitigate the risk through counselling. In this article, we evaluated the performance of the Add Health model, a multiple logistic regression model built using a novel Bayesian learning approach for predicting the risk of developing CUD in future \citep{rajapaksha2022bayesian}. The model predictors are personal level risk factors (of which two are longitudinal) measured in adolescence and/or young adulthood. The model was developed using Add Health data and its performance was evaluated here on two independent cohorts, MLS and CHDS. 
 
 The discrimination ability of the model was found to be satisfactory in datasets. The AUC was 0.73 on the CHDS data and 0.66 on MLS data. After recalibration, the E/O ratios were about 0.99 for both validation datasets. Initially, applying the model as it is to the two datasets, i.e., before recalibration, there was an under-prediction of the number of CUD cases. This under-estimation can be explained by the difference in prevalence of CUD cases between the three data cohorts with MLS and CHDS having higher prevalence of  $12.5\%$ and  $16.5\%$, respectively, as compared to the prevalence of $7.5\%$  in Add Health data.

Such variability in prevalence of CUD across the three cohorts is not unexpected. First, CHDS data was collected in New Zealand while Add Health was a study in the US. According to the 2000 World Drug Report, the annual prevalence of cannabis abuse in the population of ages 15 or above was $19.3\%$ in Oceania while it was $6.3\%$ in Americas \citep{unodc2000world}. The CHDS cohort was 30 years old and the Add Health cohort was about 18-24 years old in year 2000. Thus the higher number of CUD cases observed in CHDS as compared to Add Health is consistent with the report. Second, although both MLS and Add Health data were collected in the US, Add Health data was a nationally representative sample while MLS data were limited to one state only, specifically, Michigan. Furthermore, MLS ascertained about half of the subjects from high-risk families (Section ~\ref{paricipants}), which partially explains the higher number of CUD cases in MLS as compared to Add Health.

When there is a substantial difference in prevalence between the training and validation cohorts and that difference is due to known systematic differences such as different populations, regions, data collection procedures, or inclusion-exclusion criteria, the literature recommends correcting for this difference through a recalibration of the model intercept. \citep{moons2012risk, steyerberg2004validation, janssen2008updating}. Indeed, after the recalibration there was a substantial improvement in the E/O ratios for both validation  datasets. We also examined the robustness of the recalibration and found it to be satisfactory (see Supplementary Materials section). Note that the recalibration of intercept is not necessary if the cohorts have similar prevalences.

Although Add Health model is based on a Bayesian learning framework, an alternative may be to consider the approach used by \citet{meier2016adolescents} to identify adolescents at risk of persistent substance dependence in adulthood. This approach calculates a risk score defined as a sum of nine binary childhood and adolescent risk factors that the authors identified from the Dunedin Multidisciplinary Health and Development Study. We applied the same approach to the seven risk factors from the Add Health model. For this, we first dichotomized the six continuous variables in the model by using their medians (from the Add Health data) as cutoffs. Then we applied those cutoffs to the variables in the MLS and CHDS data. The resulting AUC values were 0.61 (0.52 – 0.69) and 0.71 (0.66 – 0.75) for MLS and CHDS datasets, respectively. These values, although lower than the corresponding values of 0.66 and 0.73, reported earlier for the Add Health model, are also reasonably good. This is partly because this simple approach inherits the benefits of the Bayesian learning approach by using the risk factors identified by it. We would like to also note that the AUC reported in \citet{meier2016adolescents} was not based on independent validation data while ours are.

The practical utility of the Add Health model as a screen for CUD can be enhanced by using it in clinical situations in which the medical or psychological evaluation and treatment of adolescents takes place. An example of this would be in the area of youth justice, in which young people who are being evaluated are generally found to have high rates of both mental health and behavioral difficulties. More generally, however, a preventive screen such as this could be used with children and young people who are identified as being from “at risk” families.

Such an approach has been shown to work effectively in a prevention science framework for understanding and evaluating the impact of identification of risk of adverse outcomes later in life, for the purposes of reducing risk of these outcomes \citep{catalano2012worldwide}. In many cases, what is being predicted in these frameworks is behavior that has not yet occurred. This may raise some ethical issues (which have been clearly elucidated by \citet{leadbeater2018ethical}). More generally, there are a wide range of ethical issues raised by research that attempts to improve prediction of mental illness, including “labelling” and stigma \citep{lawrie2019predicting}. It has been pointed out that many of these issues around early detection can be addressed to some degree by the provision of therapeutic initiatives, so that participants/patients can be assured of a meaningful and useful outcome of such detection efforts.

A limitation of our study is the relatively small sample sizes which somewhat limits our ability to accurately estimate the model's predictive power. Nonetheless, validation on two entirely different types of cohorts lends confidence toward adopting the Add Health model in practice. This model can help clinicians and counsellors in identifying adolescents/youth  early on who are at a risk of developing CUD in adulthood. Those individuals can then be helped with appropriate intervention/screening tools.

\section*{Acknowledgement}
This work was funded by the University of Texas at Dallas SPIRe seed grant. The data used in this work are from MLS and CHDS. MLS has been funded by the National Institute on Alcohol Abuse and Alcoholism (R01 AA007065 and AA025790). CHDS is supported by the Health Research Council of New Zealand (Grant 16-600: ‘The Christchurch Health and Development Study: Birth to 40 Years’). We thank the two anonymous reviewers for their thoughtful comments on the manuscript. They have led to an an improved article.


\bibliography{arxiv}

\begin{thebibliography}{30}
\expandafter\ifx\csname natexlab\endcsname\relax\def\natexlab#1{#1}\fi
\providecommand{\url}[1]{\texttt{#1}}
\providecommand{\href}[2]{#2}
\providecommand{\path}[1]{#1}
\providecommand{\DOIprefix}{doi:}
\providecommand{\ArXivprefix}{arXiv:}
\providecommand{\URLprefix}{URL: }
\providecommand{\Pubmedprefix}{pmid:}
\providecommand{\doi}[1]{\href{http://dx.doi.org/#1}{\path{#1}}}
\providecommand{\Pubmed}[1]{\href{pmid:#1}{\path{#1}}}
\providecommand{\bibinfo}[2]{#2}
\ifx\xfnm\relax \def\xfnm[#1]{\unskip,\space#1}\fi
\bibitem[{{American Psychiatric Association}(2000)}]{ap2000diagnostic}
\bibinfo{author}{{American Psychiatric Association}}, \bibinfo{year}{2000}.
\newblock \bibinfo{title}{Diagnostic and {S}tatistical {M}anual of {M}ental
  {D}isorders (4th ed.)} .
\bibitem[{Blest-Hopley et~al.(2020)Blest-Hopley, Colizzi, Giampietro and
  Bhattacharyya}]{blest2020adolescent}
\bibinfo{author}{Blest-Hopley, G.}, \bibinfo{author}{Colizzi, M.},
  \bibinfo{author}{Giampietro, V.}, \bibinfo{author}{Bhattacharyya, S.},
  \bibinfo{year}{2020}.
\newblock \bibinfo{title}{Is the adolescent brain at greater vulnerability to
  the effects of cannabis? {A} narrative review of the evidence}.
\newblock \bibinfo{journal}{Frontiers in Psychiatry} \bibinfo{volume}{11},
  \bibinfo{pages}{859}.
\bibitem[{Boden et~al.(2006)Boden, Fergusson and
  John~Horwood}]{boden2006illicit}
\bibinfo{author}{Boden, J.M.}, \bibinfo{author}{Fergusson, D.M.},
  \bibinfo{author}{John~Horwood, L.}, \bibinfo{year}{2006}.
\newblock \bibinfo{title}{Illicit drug use and dependence in a new zealand
  birth cohort}.
\newblock \bibinfo{journal}{Australian \& New Zealand Journal of Psychiatry}
  \bibinfo{volume}{40}, \bibinfo{pages}{156--163}.
\bibitem[{Caspi et~al.(2005)Caspi, Roberts and Shiner}]{caspi2005personality}
\bibinfo{author}{Caspi, A.}, \bibinfo{author}{Roberts, B.W.},
  \bibinfo{author}{Shiner, R.L.}, \bibinfo{year}{2005}.
\newblock \bibinfo{title}{Personality development: Stability and change}.
\newblock \bibinfo{journal}{Annual Review of Psychology} \bibinfo{volume}{56},
  \bibinfo{pages}{453--484}.
\bibitem[{Casswell et~al.(1991)Casswell, Stewart, Connolly and
  Silva}]{casswell1991longitudinal}
\bibinfo{author}{Casswell, S.}, \bibinfo{author}{Stewart, J.},
  \bibinfo{author}{Connolly, G.}, \bibinfo{author}{Silva, P.},
  \bibinfo{year}{1991}.
\newblock \bibinfo{title}{A longitudinal study of {N}ew {Z}ealand children's
  experience with alcohol}.
\newblock \bibinfo{journal}{British Journal of Addiction} \bibinfo{volume}{86},
  \bibinfo{pages}{277--285}.
\bibitem[{Catalano et~al.(2012)Catalano, Fagan, Gavin, Greenberg, Irwin, Ross
  et~al.}]{catalano2012worldwide}
\bibinfo{author}{Catalano, R.F.}, \bibinfo{author}{Fagan, A.A.},
  \bibinfo{author}{Gavin, L.E.}, \bibinfo{author}{Greenberg, M.T.},
  \bibinfo{author}{Irwin, C.E.}, \bibinfo{author}{Ross, D.A.}, et~al.,
  \bibinfo{year}{2012}.
\newblock \bibinfo{title}{Worldwide application of prevention science in
  adolescent health}.
\newblock \bibinfo{journal}{The Lancet} \bibinfo{volume}{379},
  \bibinfo{pages}{1653--1664}.
\bibitem[{Costa and McCrae(1992)}]{mccrae1992revised}
\bibinfo{author}{Costa, P.T.}, \bibinfo{author}{McCrae, R.R.},
  \bibinfo{year}{1992}.
\newblock \bibinfo{title}{Normal personality assessment in clinical practice:
  The {NEO} personality inventory.}
\newblock \bibinfo{journal}{Psychological Assessment} \bibinfo{volume}{4},
  \bibinfo{pages}{5--13}.
\bibitem[{Damian et~al.(2019)Damian, Spengler, Sutu and
  Roberts}]{damian2019sixteen}
\bibinfo{author}{Damian, R.I.}, \bibinfo{author}{Spengler, M.},
  \bibinfo{author}{Sutu, A.}, \bibinfo{author}{Roberts, B.W.},
  \bibinfo{year}{2019}.
\newblock \bibinfo{title}{Sixteen going on sixty-six: A longitudinal study of
  personality stability and change across 50 years.}
\newblock \bibinfo{journal}{Journal of Personality and Social Psychology}
  \bibinfo{volume}{117}, \bibinfo{pages}{674}.
\bibitem[{{Harris, K.M.and Halpern, C.T.and Whitsel, E.and Hussey, J.and Tabor,
  J.and Entzel, P.and Udry, J.R.}(2009)}]{harriss2009}
\bibinfo{author}{{Harris, K.M.and Halpern, C.T.and Whitsel, E.and Hussey, J.and
  Tabor, J.and Entzel, P.and Udry, J.R.}}, \bibinfo{year}{2009}.
\newblock \bibinfo{title}{The {N}ational {L}ongitudinal {S}tudy of {A}dolescent
  to {A}dult {H}ealth: Research {D}esign}.
\newblock
  \bibinfo{howpublished}{\url{https://addhealth.cpc.unc.edu/documentation/study-design/}}.
\newblock \bibinfo{note}{(Accessed Jan 20, 2023)}.
\bibitem[{Janssen et~al.(2008)Janssen, Moons, Kalkman, Grobbee and
  Vergouwe}]{janssen2008updating}
\bibinfo{author}{Janssen, K.}, \bibinfo{author}{Moons, K.},
  \bibinfo{author}{Kalkman, C.}, \bibinfo{author}{Grobbee, D.},
  \bibinfo{author}{Vergouwe, Y.}, \bibinfo{year}{2008}.
\newblock \bibinfo{title}{Updating methods improved the performance of a
  clinical prediction model in new patients}.
\newblock \bibinfo{journal}{Journal of Clinical Epidemiology}
  \bibinfo{volume}{61}, \bibinfo{pages}{76--86}.
\bibitem[{Lawrie et~al.(2019)Lawrie, Fletcher-Watson, Whalley and
  McIntosh}]{lawrie2019predicting}
\bibinfo{author}{Lawrie, S.M.}, \bibinfo{author}{Fletcher-Watson, S.},
  \bibinfo{author}{Whalley, H.C.}, \bibinfo{author}{McIntosh, A.M.},
  \bibinfo{year}{2019}.
\newblock \bibinfo{title}{Predicting major mental illness: {E}thical and
  practical considerations}.
\newblock \bibinfo{journal}{BJPsych Open} \bibinfo{volume}{5},
  \bibinfo{pages}{e30}.
\bibitem[{Leadbeater et~al.(2018)Leadbeater, Dishion, Sandler, Bradshaw, Dodge,
  Gottfredson et~al.}]{leadbeater2018ethical}
\bibinfo{author}{Leadbeater, B.J.}, \bibinfo{author}{Dishion, T.},
  \bibinfo{author}{Sandler, I.}, \bibinfo{author}{Bradshaw, C.P.},
  \bibinfo{author}{Dodge, K.}, \bibinfo{author}{Gottfredson, D.}, et~al.,
  \bibinfo{year}{2018}.
\newblock \bibinfo{title}{Ethical challenges in promoting the implementation of
  preventive interventions: Report of the {SPR} task force}.
\newblock \bibinfo{journal}{Prevention Science} \bibinfo{volume}{19},
  \bibinfo{pages}{853--865}.
\bibitem[{Meier et~al.(2016)Meier, Hall, Caspi, Belsky, Cerd{\'a}, Harrington,
  Houts, Poulton and Moffitt}]{meier2016adolescents}
\bibinfo{author}{Meier, M.H.}, \bibinfo{author}{Hall, W.},
  \bibinfo{author}{Caspi, A.}, \bibinfo{author}{Belsky, D.W.},
  \bibinfo{author}{Cerd{\'a}, M.}, \bibinfo{author}{Harrington, H.},
  \bibinfo{author}{Houts, R.}, \bibinfo{author}{Poulton, R.},
  \bibinfo{author}{Moffitt, T.E.}, \bibinfo{year}{2016}.
\newblock \bibinfo{title}{Which adolescents develop persistent substance
  dependence in adulthood? {U}sing population-representative longitudinal data
  to inform universal risk assessment}.
\newblock \bibinfo{journal}{Psychological Medicine} \bibinfo{volume}{46},
  \bibinfo{pages}{877--889}.
\bibitem[{Moons et~al.(2012)Moons, Kengne, Grobbee, Royston, Vergouwe, Altman
  and Woodward}]{moons2012risk}
\bibinfo{author}{Moons, K.G.M.}, \bibinfo{author}{Kengne, A.P.},
  \bibinfo{author}{Grobbee, D.E.}, \bibinfo{author}{Royston, P.},
  \bibinfo{author}{Vergouwe, Y.}, \bibinfo{author}{Altman, D.G.},
  \bibinfo{author}{Woodward, M.}, \bibinfo{year}{2012}.
\newblock \bibinfo{title}{Risk prediction models: {II}. {E}xternal validation,
  model updating, and impact assessment}.
\newblock \bibinfo{journal}{Heart} \bibinfo{volume}{98},
  \bibinfo{pages}{691--698}.
\bibitem[{{NCDAS}(2022a)}]{ncdas_2022}
\bibinfo{author}{{NCDAS}}, \bibinfo{year}{2022}a.
\newblock \bibinfo{title}{Drug abuse statistics}.
\newblock \bibinfo{howpublished}{\url{https://drugabusestatistics.org/}}.
\newblock \bibinfo{note}{(Accessed June 02, 2022)}.
\bibitem[{{NCDAS}(2022b)}]{ncdas_2022_youth}
\bibinfo{author}{{NCDAS}}, \bibinfo{year}{2022}b.
\newblock \bibinfo{title}{Drug use among youth: Facts \& statistics}.
\newblock
  \bibinfo{howpublished}{\url{https://drugabusestatistics.org/teen-drug-use/}}.
\newblock \bibinfo{note}{(Accessed June 02, 2022)}.
\bibitem[{Poulton et~al.(2020)Poulton, Robertson, Boden, Horwood, Theodore,
  Potiki and Ambler}]{poulton2020patterns}
\bibinfo{author}{Poulton, R.}, \bibinfo{author}{Robertson, K.},
  \bibinfo{author}{Boden, J.}, \bibinfo{author}{Horwood, J.},
  \bibinfo{author}{Theodore, R.}, \bibinfo{author}{Potiki, T.},
  \bibinfo{author}{Ambler, A.}, \bibinfo{year}{2020}.
\newblock \bibinfo{title}{Patterns of recreational cannabis use in
  {A}otearoa-{N}ew {Z}ealand and their consequences: {E}vidence to inform
  voters in the 2020 referendum}.
\newblock \bibinfo{journal}{Journal of the Royal Society of New Zealand}
  \bibinfo{volume}{50}, \bibinfo{pages}{348--365}.
\bibitem[{{R Core Team}(2022)}]{Rstudio}
\bibinfo{author}{{R Core Team}}, \bibinfo{year}{2022}.
\newblock \bibinfo{title}{R: A Language and Environment for Statistical
  Computing}.
\newblock \bibinfo{organization}{R Foundation for Statistical Computing}.
  \bibinfo{address}{Vienna, Austria}.
\newblock \URLprefix \url{https://www.R-project.org/}.
\bibitem[{Rajapaksha et~al.(2022)Rajapaksha, Filbey, Biswas and
  Choudhary}]{rajapaksha2022bayesian}
\bibinfo{author}{Rajapaksha, R.M.D.S.}, \bibinfo{author}{Filbey, F.},
  \bibinfo{author}{Biswas, S.}, \bibinfo{author}{Choudhary, P.},
  \bibinfo{year}{2022}.
\newblock \bibinfo{title}{A {B}ayesian learning model to predict the risk for
  cannabis use disorder}.
\newblock \bibinfo{journal}{Drug and Alcohol Dependence} \bibinfo{volume}{236},
  \bibinfo{pages}{109476}.
\bibitem[{Robin et~al.(2011)Robin, Turck, Hainard, Tiberti, Lisacek, Sanchez
  and Müller}]{proc}
\bibinfo{author}{Robin, X.}, \bibinfo{author}{Turck, N.},
  \bibinfo{author}{Hainard, A.}, \bibinfo{author}{Tiberti, N.},
  \bibinfo{author}{Lisacek, F.}, \bibinfo{author}{Sanchez, J.C.},
  \bibinfo{author}{Müller, M.}, \bibinfo{year}{2011}.
\newblock \bibinfo{title}{proc: an open-source package for r and s+ to analyze
  and compare roc curves}.
\newblock \bibinfo{journal}{BMC Bioinformatics} \bibinfo{volume}{12},
  \bibinfo{pages}{77}.
\bibitem[{Robins et~al.(1981)Robins, Helzer, Croughan and
  Ratcliff}]{robins1995diagnostic}
\bibinfo{author}{Robins, L.N.}, \bibinfo{author}{Helzer, J.E.},
  \bibinfo{author}{Croughan, J.}, \bibinfo{author}{Ratcliff, K.S.},
  \bibinfo{year}{1981}.
\newblock \bibinfo{title}{National institute of mental health diagnostic
  interview schedule: Its history, characteristics, and validity.}
\newblock \bibinfo{journal}{Archives of General Psychiatry}
  \bibinfo{volume}{38}, \bibinfo{pages}{381--389}.
\bibitem[{{SAMHSA}(2021)}]{abuse2020key}
\bibinfo{author}{{SAMHSA}}, \bibinfo{year}{2021}.
\newblock \bibinfo{title}{Key substance use and mental health indicators in the
  united states: Results from the 2020 national survey on drug use and health}.
\newblock \bibinfo{howpublished}{\url{https://www.samhsa.gov/data/}}.
\newblock \bibinfo{note}{(Accessed June 02, 2022)}.
\bibitem[{Silins et~al.(2014)Silins, Horwood, Patton, Fergusson, Olsson,
  Hutchinson, Spry, Toumbourou, Degenhardt, Swift et~al.}]{silins2014young}
\bibinfo{author}{Silins, E.}, \bibinfo{author}{Horwood, L.J.},
  \bibinfo{author}{Patton, G.C.}, \bibinfo{author}{Fergusson, D.M.},
  \bibinfo{author}{Olsson, C.A.}, \bibinfo{author}{Hutchinson, D.M.},
  \bibinfo{author}{Spry, E.}, \bibinfo{author}{Toumbourou, J.W.},
  \bibinfo{author}{Degenhardt, L.}, \bibinfo{author}{Swift, W.}, et~al.,
  \bibinfo{year}{2014}.
\newblock \bibinfo{title}{Young adult sequelae of adolescent cannabis use: an
  integrative analysis}.
\newblock \bibinfo{journal}{The Lancet Psychiatry} \bibinfo{volume}{1},
  \bibinfo{pages}{286--293}.
\bibitem[{Steyerberg et~al.(2004)Steyerberg, Borsboom, van Houwelingen,
  Eijkemans and Habbema}]{steyerberg2004validation}
\bibinfo{author}{Steyerberg, E.W.}, \bibinfo{author}{Borsboom, G.J.J.M.},
  \bibinfo{author}{van Houwelingen, H.C.}, \bibinfo{author}{Eijkemans, M.J.C.},
  \bibinfo{author}{Habbema, J.D.F.}, \bibinfo{year}{2004}.
\newblock \bibinfo{title}{Validation and updating of predictive logistic
  regression models: A study on sample size and shrinkage}.
\newblock \bibinfo{journal}{Statistics in Medicine} \bibinfo{volume}{23},
  \bibinfo{pages}{2567--2586}.
\bibitem[{Taylor et~al.(2019)Taylor, Cousijn and
  Filbey}]{taylor2019determining}
\bibinfo{author}{Taylor, M.}, \bibinfo{author}{Cousijn, J.},
  \bibinfo{author}{Filbey, F.}, \bibinfo{year}{2019}.
\newblock \bibinfo{title}{Determining risks for cannabis use disorder in the
  face of changing legal policies}.
\newblock \bibinfo{journal}{Current Addiction Reports} \bibinfo{volume}{6},
  \bibinfo{pages}{466--477}.
\bibitem[{Wittchen et~al.(2007)Wittchen, Fr{\"o}hlich, Behrendt, G{\"u}nther,
  Rehm, Zimmermann, Lieb and Perkonigg}]{wittchen2007cannabis}
\bibinfo{author}{Wittchen, H.U.}, \bibinfo{author}{Fr{\"o}hlich, C.},
  \bibinfo{author}{Behrendt, S.}, \bibinfo{author}{G{\"u}nther, A.},
  \bibinfo{author}{Rehm, J.}, \bibinfo{author}{Zimmermann, P.},
  \bibinfo{author}{Lieb, R.}, \bibinfo{author}{Perkonigg, A.},
  \bibinfo{year}{2007}.
\newblock \bibinfo{title}{Cannabis use and cannabis use disorders and their
  relationship to mental disorders: {A} 10-year prospective-longitudinal
  community study in adolescents}.
\newblock \bibinfo{journal}{Drug and Alcohol Dependence} \bibinfo{volume}{88},
  \bibinfo{pages}{S60--S70}.
\bibitem[{{World {D}rug {R}eport}(2001)}]{unodc2000world}
\bibinfo{author}{{World {D}rug {R}eport}}, \bibinfo{year}{2001}.
\newblock \bibinfo{title}{United {N}ations publications, {S}ales {N}o
  {GV.E}.00.0.10}.
\newblock
  \bibinfo{howpublished}{\url{https://www.unodc.org/unodc/en/data-and-analysis/WDR-2000.html}}.
\newblock \bibinfo{note}{(Accessed June 02, 2022)}.
\bibitem[{{World Health Organization}(1994)}]{CIDI}
\bibinfo{author}{{World Health Organization}}, \bibinfo{year}{1994}.
\newblock \bibinfo{title}{Composite {I}nternational {D}iagnostic {I}nterview
  ({C}{I}{D}{I}) researcher's manual (Version 1.1)}.
\bibitem[{Zucker et~al.(1996)Zucker, Ellis, Fitzgerald, Bingham and
  Sanford}]{zucker1996other}
\bibinfo{author}{Zucker, R.A.}, \bibinfo{author}{Ellis, D.A.},
  \bibinfo{author}{Fitzgerald, H.E.}, \bibinfo{author}{Bingham, C.R.},
  \bibinfo{author}{Sanford, K.}, \bibinfo{year}{1996}.
\newblock \bibinfo{title}{Other evidence for at least two alcoholisms {II}:
  Life course variation in antisociality and heterogeneity of alcoholic
  outcome}.
\newblock \bibinfo{journal}{Development and Psychopathology}
  \bibinfo{volume}{8}, \bibinfo{pages}{831--848}.
\bibitem[{Zucker et~al.(2000)Zucker, Fitzgerald, Refior, Puttler, Pallas and
  Ellis}]{zucker2002clinical}
\bibinfo{author}{Zucker, R.A.}, \bibinfo{author}{Fitzgerald, H.E.},
  \bibinfo{author}{Refior, S.K.}, \bibinfo{author}{Puttler, L.I.},
  \bibinfo{author}{Pallas, D.M.}, \bibinfo{author}{Ellis, D.A.},
  \bibinfo{year}{2000}.
\newblock \bibinfo{title}{The clinical and social ecology of childhood for
  children of alcoholics: Description of a study and implications for a
  differentiated social policy}, in: \bibinfo{editor}{Fitzgerald, H.E.},
  \bibinfo{editor}{Lesterand, B.M.}, \bibinfo{editor}{Zuckerman, B.S.} (Eds.),
  \bibinfo{booktitle}{Children of Addiction}. \bibinfo{publisher}{Routledge},
  pp. \bibinfo{pages}{125--158}.

\end{thebibliography}


\begin{thebibliography}{13}
\expandafter\ifx\csname natexlab\endcsname\relax\def\natexlab#1{#1}\fi
\providecommand{\url}[1]{\texttt{#1}}
\providecommand{\href}[2]{#2}
\providecommand{\path}[1]{#1}
\providecommand{\DOIprefix}{doi:}
\providecommand{\ArXivprefix}{arXiv:}
\providecommand{\URLprefix}{URL: }
\providecommand{\Pubmedprefix}{pmid:}
\providecommand{\doi}[1]{\href{http://dx.doi.org/#1}{\path{#1}}}
\providecommand{\Pubmed}[1]{\href{pmid:#1}{\path{#1}}}
\providecommand{\bibinfo}[2]{#2}
\ifx\xfnm\relax \def\xfnm[#1]{\unskip,\space#1}\fi
\bibitem[{Bavo et~al.(2016)Bavo, Daan, Ben, Ewout and Yvonne}]{calibrartion}
\bibinfo{author}{Bavo, D.C.}, \bibinfo{author}{Daan, N.}, \bibinfo{author}{Ben,
  V.C.}, \bibinfo{author}{Ewout, S.}, \bibinfo{author}{Yvonne, V.},
  \bibinfo{year}{2016}.
\newblock \bibinfo{title}{CalibrationCurves: Calibration performance}.
\newblock \bibinfo{note}{R package version 0.1.2}.
\bibitem[{Cloninger(1987)}]{cloninger1987systematic}
\bibinfo{author}{Cloninger, C.R.}, \bibinfo{year}{1987}.
\newblock \bibinfo{title}{A systematic method for clinical description and
  classification of personality variants: A proposal}.
\newblock \bibinfo{journal}{Archives of General Psychiatry}
  \bibinfo{volume}{44}, \bibinfo{pages}{573--588}.
\bibitem[{Conners(1969)}]{conners1969teacher}
\bibinfo{author}{Conners, C.K.}, \bibinfo{year}{1969}.
\newblock \bibinfo{title}{A teacher rating scale for use in drug studies with
  children}.
\newblock \bibinfo{journal}{American Journal of Psychiatry}
  \bibinfo{volume}{126}, \bibinfo{pages}{884--888}.
\bibitem[{Conners(1970)}]{conners1970symptom}
\bibinfo{author}{Conners, C.K.}, \bibinfo{year}{1970}.
\newblock \bibinfo{title}{Symptom patterns in hyperkinetic, neurotic, and
  normal children}.
\newblock \bibinfo{journal}{Child Development} \bibinfo{volume}{41},
  \bibinfo{pages}{667--682}.
\bibitem[{Costa and McCrae(1992)}]{mccrae1992revised}
\bibinfo{author}{Costa, P.T.}, \bibinfo{author}{McCrae, R.R.},
  \bibinfo{year}{1992}.
\newblock \bibinfo{title}{Normal personality assessment in clinical practice:
  The {NEO} personality inventory.}
\newblock \bibinfo{journal}{Psychological Assessment} \bibinfo{volume}{4},
  \bibinfo{pages}{5--13}.
\bibitem[{Costello et~al.(1985)Costello, Edelbrock and
  Costello}]{costello1985validity}
\bibinfo{author}{Costello, E.J.}, \bibinfo{author}{Edelbrock, C.S.},
  \bibinfo{author}{Costello, A.J.}, \bibinfo{year}{1985}.
\newblock \bibinfo{title}{Validity of the {NIMH} diagnostic interview schedule
  for children: A comparison between psychiatric and pediatric referrals}.
\newblock \bibinfo{journal}{Journal of Abnormal Child Psychology}
  \bibinfo{volume}{13}, \bibinfo{pages}{579--595}.
\bibitem[{Eysenck and Sybil(1964)}]{eysenck1964improved}
\bibinfo{author}{Eysenck, H.J.}, \bibinfo{author}{Sybil, B.},
  \bibinfo{year}{1964}.
\newblock \bibinfo{title}{An improved short questionnaire for the measurement
  of extraversion and neuroticism.}
\newblock \bibinfo{journal}{Life Sciences} \bibinfo{volume}{3},
  \bibinfo{pages}{1103--1109}.
\bibitem[{Janssen et~al.(2008)Janssen, Moons, Kalkman, Grobbee and
  Vergouwe}]{janssen2008updating}
\bibinfo{author}{Janssen, K.}, \bibinfo{author}{Moons, K.},
  \bibinfo{author}{Kalkman, C.}, \bibinfo{author}{Grobbee, D.},
  \bibinfo{author}{Vergouwe, Y.}, \bibinfo{year}{2008}.
\newblock \bibinfo{title}{Updating methods improved the performance of a
  clinical prediction model in new patients}.
\newblock \bibinfo{journal}{Journal of Clinical Epidemiology}
  \bibinfo{volume}{61}, \bibinfo{pages}{76--86}.
\bibitem[{Moons et~al.(2012)Moons, Kengne, Grobbee, Royston, Vergouwe, Altman
  and Woodward}]{moons2012risk}
\bibinfo{author}{Moons, K.G.M.}, \bibinfo{author}{Kengne, A.P.},
  \bibinfo{author}{Grobbee, D.E.}, \bibinfo{author}{Royston, P.},
  \bibinfo{author}{Vergouwe, Y.}, \bibinfo{author}{Altman, D.G.},
  \bibinfo{author}{Woodward, M.}, \bibinfo{year}{2012}.
\newblock \bibinfo{title}{Risk prediction models: {II}. {E}xternal validation,
  model updating, and impact assessment}.
\newblock \bibinfo{journal}{Heart} \bibinfo{volume}{98},
  \bibinfo{pages}{691--698}.
\bibitem[{Rutter et~al.(1970)Rutter, Tizard and Whitmore}]{rutter1970education}
\bibinfo{author}{Rutter, M.}, \bibinfo{author}{Tizard, J.},
  \bibinfo{author}{Whitmore, K.}, \bibinfo{year}{1970}.
\newblock \bibinfo{title}{Education, Health and Behaviour: Psychological and
  Medical Study of Childhood Development, first ed}.
\newblock \bibinfo{publisher}{Longman Group}, \bibinfo{address}{London}.
\bibitem[{Sibley(2012)}]{sibley2012mini}
\bibinfo{author}{Sibley, C.G.}, \bibinfo{year}{2012}.
\newblock \bibinfo{title}{The mini-{I}{P}{I}{P}6: Item response theory analysis
  of a short measure of the big-six factors of personality in {N}ew {Z}ealand.}
\newblock \bibinfo{journal}{New Zealand Journal of Psychology}
  \bibinfo{volume}{41}.
\bibitem[{Steyerberg et~al.(2004)Steyerberg, Borsboom, van Houwelingen,
  Eijkemans and Habbema}]{steyerberg2004validation}
\bibinfo{author}{Steyerberg, E.W.}, \bibinfo{author}{Borsboom, G.J.J.M.},
  \bibinfo{author}{van Houwelingen, H.C.}, \bibinfo{author}{Eijkemans, M.J.C.},
  \bibinfo{author}{Habbema, J.D.F.}, \bibinfo{year}{2004}.
\newblock \bibinfo{title}{Validation and updating of predictive logistic
  regression models: A study on sample size and shrinkage}.
\newblock \bibinfo{journal}{Statistics in Medicine} \bibinfo{volume}{23},
  \bibinfo{pages}{2567--2586}.
\bibitem[{Van~Calster et~al.(2019)Van~Calster, McLernon, Van~Smeden, Wynants
  and Steyerberg}]{van2019calibration}
\bibinfo{author}{Van~Calster, B.}, \bibinfo{author}{McLernon, D.J.},
  \bibinfo{author}{Van~Smeden, M.}, \bibinfo{author}{Wynants, L.},
  \bibinfo{author}{Steyerberg, E.W.}, \bibinfo{year}{2019}.
\newblock \bibinfo{title}{Calibration: The {A}chilles heel of predictive
  analytics}.
\newblock \bibinfo{journal}{BMC Medicine} \bibinfo{volume}{17},
  \bibinfo{pages}{1--7}.

\end{thebibliography}


\newpage 
   
\begin{longtable}[htp]{p{1.2in}p{1.2in}p{1.6in}p{2.3in}}
\caption{Comparison of Predictors in Add Health, MLS, and CHDS data}\\
\hline

\label{VAr comparison}    
Predictor	 & Add Health & 	MLS	 & CHDS \\ \hline
Biological Sex&	Available & Available & 	Available \\ 
Conscientiousness scale, Neuroticism scale and Openness scale & Personality questionnaire where each scale is measured by 4 questions. & Neuroticism, openness and conscientiousness was obtained from 60-item NEO-FFI. Each personality domain was measured by 12 questions. Measure considered at wave 5 (ages 12 -- 19) .  &Measures neuroticism (age 14) using a short form version of the Eysenck Personality Inventory, openness using corresponding items from Tri-dimensional Personality Questionnaire (age 16) and Conscientiousness was
using a short-form personality instrument called  Mini-IPIP6 (age 40).\\

Peer cannabis use & Number of best friends (out of 3 best friends) who use cannabis at least once a month.	& Annually administered peer behavior profile (during ages 11 -- 17) asked "how many of the friends you hang out with most of the
time get high on drugs once a month or more often".	& At ages 15 and 16 ("best friend use cannabis", "other close friends use cannabis" in last year); at ages 18, 21 and 25 ("How many of your Male/female friends Use marijuana or other drugs").   \\

Delinquency&
Number of times involved in delinquent activities. A higher value implies a greater involvement. &	A subset of annually administered (during ages 11 -- 17) anti-social behavior (ASB) questionnaire.  We closely matched the questions with Add Health delinquency questionnaire and selected 23 questions from ASB. & Measured delinquency from ages 7 to 16. From ages 7-10, delinquency scores are derived combining Rutter  and Conners  parent and teacher questionnaires. From age 10-16 the final score is a combination of parent, teacher and self reports where self report is a  questionnaire based on items derived from the Diagnostic Interview Schedule for Children.\\
ACE&  Number of adverse childhood experiences that occurred before age 18.  & 	Out of the 6 questions that make up ACE in Add Health, we found 5 in
MLS. These include sexual Abuse, physical Abuse, parental separation/divorce or death, parental incarceration, and household substance abuse. The age at event occurrence was taken into account to make sure event happened before age 18.	& We found the same 5
questions in CHDS corresponding to the questions that make up ACE in Add Health. The age at event occurrence was taken into account to make sure event happened before age 18. \\ \hline

\end{longtable}
 \setlength{\tabcolsep}{6pt}  
 
\clearpage

\begin{table} [h]

     \centering   
      \caption{Summary of predictors in MLS data:  mean (standard deviation) for continuous predictors; Delinquency: mean (standard deviation) of average score over the waves in which it was measured, Peer cannabis use: mean (standard deviation) of random effects over all subjects}
      \label{summary_cross_MLS}
            \begin{tabular}{l l l l} 
 \hline
 Predictor	&\makecell{Total\\ ($n$ = 424)}  &	\makecell{Cases\\($n$ = 53)}&		\makecell{Controls\\ ($n$ = 371)}\\ \hline
 Male \% & 70.28 & 77.36 &69.27 \\ 
ACE scale  &0.18 (0.22)& 0.20 (0.20)&  0.18 (0.22)\\
Neuroticism scale &0.52 (0.13)& 0.55 (0.11) &0.51 (0.13) \\ 
Conscientiousness scale&  0.69 (0.12)& 0.64 (0.10) & 0.69 (0.12)\\ 
Openness scale &0.62 (0.10)&0.62 (0.10)& 0.62 (0.10)\\ 
Delinquency & 0.08 (0.07) & 0.11 (0.08) & 0.07 (0.06) \\ 
Peer cannabis use & 0.12 (0.10) & 0.16 (0.11) & 0.12 (0.09)\\  \hline
 
              \end{tabular}
            \end{table}

\begin{table} [h]

     \centering   
      \caption{Summary of predictors in CHDS data: mean (standard deviation) for continuous predictors; Delinquency: mean (standard deviation) of average score over the waves in which it was measured, Peer cannabis use: mean (standard deviation) of random effects over all subjects}
      \label{summary_cross_CHDS}
            \begin{tabular}{l l l l} 
 \hline
  Predictor	&\makecell{Total\\ ($n$ = 637)}  &	\makecell{Cases\\($n$ = 105)}&		\makecell{Controls\\ ($n$ = 532)}\\ \hline
 Male \% & 50.07 & 70.48&46.05\\ 
ACE scale  &0.35 (0.20)& 0.41 (0.23)&  0.34(0.19)\\ 
Neuroticism scale &0.14 (0.13)& 0.16 (0.12) &0.14 (0.13) \\ 
Conscientiousness scale&  0.48 (0.14)& 0.46 (0.13) & 0.49 (0.14)\\ 
Openness scale &0.61 (0.16)&0.68 (0.16)& 0.60 (0.16)\\
Delinquency & 0.09 (0.06) & 0.12 (0.06) & 0.08 (0.06) \\ 
Peer cannabis use & 0.16 (0.17) & 0.21 (0.19) & 0.14 (0.16)\\  \hline
 
              \end{tabular}
            \end{table}
\clearpage

 \setlength{\tabcolsep}{20pt}           
 \begin{table} [!htpb]

     \centering   
      \caption{Expected (E) and observed (O) number of CUD cases for MLS data}
      \label{quantiles_MLS}
            \begin{tabular}{l l l l l} 
 \hline
	 & $n$ & 	E	 & O &	E/O\\ \hline
{\bf Risk quantile groups}	&\multicolumn{4}{l}{}\\ 
Group 1&	85&	5.218&	4&	1.304\\ 
Group 2	&85&	7.480&	9&	0.831\\ 
Group 3	&84&	9.033&	8&	1.129\\ 
Group 4&	85&	11.844&	14	&0.846\\ 
Group 5	&85	&19.411	&18&	1.078\\ 
{\bf Biological Sex}	&\multicolumn{4}{l}{}\\ 		 
Male&	298	& 40.082 &41	&0.978\\ 
Female&	126	&12.905&	12&	1.075\\ 
{\bf ACE scale} &	\multicolumn{4}{l}{}	\\			 
Below median&	246	&26.627&	30&	0.888\\ 
Above median&	178&	26.359&	23&	1.146\\ 
{\bf Conscientiousness scale} &	\multicolumn{4}{l}{}	\\ 		 
Below median&	232&	34.146&	35&	0.976\\
Above median	&192	&18.840&	18&	1.047\\
{\bf Neuroticism scale} &\multicolumn{4}{l}{} \\ 		 
Below median&	221	&22.093&	26	&0.850\\ 
Above median&	203	&30.893&	27	&1.144\\ 
{\bf Openness scale}	& \multicolumn{4}{l}{} \\ 		 
Below median&	241	&27.336&	32&	0.854\\ 
Above median	&183	&25.650&	21	&1.222\\ 
{\bf Delinquency scale}	& \multicolumn{4}{l}{} \\ 			 
Below median&	214&	19.690&	18&	1.094\\ 
Above median	&210	&33.296&	35	&0.951\\ 
{\bf Peer cannabis use} &	 \multicolumn{4}{l}{} \\ 			 
Below median&	212&	22.968&	19	&1.209\\ 
Above median	&212&	30.018	&34	&0.883\\ \hline

\end{tabular}
            \end{table}

\begin{table} [htp]

     \centering   
      \caption{Expected (E) and observed (O) number of CUD cases for CHDS data}
      \label{quantiles_CHDS}
            \begin{tabular}{l l l l l} 
 \hline
 & $n$ & 	E	 & O &	E/O\\ \hline
{\bf Risk quantile groups} &	\multicolumn{4}{l}{}	\\
Group 1&	128&	9.074&	3&	3.025\\ 
Group 2	&127&13.865&	11&	1.260\\ 
Group 3	&127&	18.640&	17&	1.096\\
Group 4&	127&	24.171 &	35	&0.691\\ 
Group 5	&128	&38.721	&39&	0.993\\ 
{\bf Biological Sex}	&\multicolumn{4}{l}{}\\ 	 
Male&	319	&57.827	&74	&0.781\\ 
Female&	318	&46.645&	31&	1.505\\ 
{\bf ACE scale} &	\multicolumn{4}{l}{}	\\ 		 
Below median&	334	&45.082&	42&	1.073\\ 
Above median&	303&	59.390&	63&	0.943\\ 
{\bf Conscientiousness scale} &	\multicolumn{4}{l}{}	\\ 	 
Below median&	411&	75.885&	73&	1.040\\ 
Above median	&226	&28.587&	32&	0.893\\ 
{\bf Neuroticism scale} &\multicolumn{4}{l}{} \\ 	 
Below median&	337	&46.455&	45	&1.032\\ 
Above median&	300	&58.017&	60	&0.967\\ 
{\bf Openness scale}	& \multicolumn{4}{l}{} \\ 		 
Below median&	338	&43.251&	38&	1.138\\ 
Above median	&299	&61.221&	67	&0.914\\ 
{\bf Delinquency scale}	& \multicolumn{4}{l}{} \\ 		 
Below median&	319&	36.513&	31&	1.178\\ 
Above median	&318	&67.958&	74	&0.918\\ 
{\bf Peer cannabis use} &	 \multicolumn{4}{l}{} \\ 		 
Below median&	326&	43.964&	42	&1.047\\ 
Above median	&311&	60.508	&63	&0.960\\ \hline

              \end{tabular}
            \end{table}


 


\newpage

\end{document}


\setcounter{page}{22}
\maketitle

\section{Details on Measures Used in Model Validation}
We tried to match the seven risk factors in the Add Health model with relevant measures from MLS and CHDS to the extent possible as described below

\subsection{Peer Cannabis Use}
In Add Health model, this variable was defined based on the Add Health question "Of your 3 best friends, how many use marijuana at least once a month?" In MLS, peer behavior profile, an unpublished questionnaire annually administered during the ages 11-17 (7 waves) asked "how many of the friends you hang out with most of the time get high on drugs once a month or more often". The answers were on a Likert scale ranging from 1-5, with 5 being the highest category. The answer was divided by 5 to convert to the 0-1 scale required by the model.  

In CHDS, at ages 15 and 16, two questions were asked regarding the use of cannabis by best friend and close friends in the last year ("does the best friend use cannabis" , "do other close friends use cannabis") to which the answers were yes (1) or no (0). Also, at ages 18, 21, and 25, two questions were asked about the current cannabis use by male and female friends ("How many of your male/female friends use marijuana or other drugs") The answers were on a Likert scale of 1 to 3, where 3 referred to "most friends". At each age, the average of the two answers were taken and converted to a 0-1 scale.

\subsection{Delinquency}
\label{delinq}
Add Health measured the number of times a participant was involved in various types of delinquent activities. A higher value implied a greater involvement. The corresponding MLS measure was obtained through a subset of questions included in annually administered anti-social behavior (ASB) checklist (unpublished questionnaire) during the ages 11-17. We closely matched the questions with Add Health delinquency questionnaire and selected 23 questions from ASB. Each question asked "how often you've done this activity". The response was on a Likert scale ranging from 0-3; 3 being more frequent behavior. When multiple questions in ASB matched a single question in Add Health delinquency questionnaire we used the one that was most frequent for a participant. The final score for each participant was the sum of the responses for non-missing questions divided by the number of non-missing questions. To match with the scale of Add Health model (0 to 1), we further divided this score by 3.

The CHDS measured delinquency from ages 7 to 16. For ages 7-9, delinquency scores were derived from both parent and teacher reports using an instrument combining Rutter \citep{rutter1970education} and Conners \citep{conners1969teacher,conners1970symptom} parent and teacher questionnaires. From ages 10 to 16, the final score was a combination of parent, teacher, and self report, where self report is a  questionnaire based on items derived from the Diagnostic Interview Schedule for Children \citep{costello1985validity}. An unusual feature is that the minimum possible score for the questionnaire was not 0. Therefore, to convert the final delinquency score to 0-1 scale at a given age, we first subtracted minimum possible score for the questionnaire (for that age) from the observed score and then divided by the maximum possible score.

\subsection{Neuroticism, Openness, and Conscientiousness}
Add Health measured personality with a questionnaire where each trait was measured by 4 questions. Personality traits in MLS were obtained from 60-item  NEO Five Factor Inventory (NEO-FFI) \citep{mccrae1992revised} wherein each trait is measured by 12 questions. We used the measure from wave 5 wherein the participants were around age 14 to 19 since it had the most non-missing values. Each personality trait had final scores between 0-60, which we divided by 60.

CHDS measured neuroticism at age 14 using a short form version of the Eysenck personality inventory \citep{eysenck1964improved} and openness at age 16 using corresponding items from tri-dimensional personality questionnaire \citep{cloninger1987systematic}. Conscientiousness was measured at age 40 using a short-form personality instrument \citep{sibley2012mini}. Similar to delinquency, the minimum possible score for the questionnaires was not 0. Therefore, the observed scores were scaled between 0-1 in the same manner as described for Section \ref{delinq}.

\subsection{Adverse Childhood Experiences (ACE)}
Add Health measured the number of ACE that occurred before age 18. Out of the 6 questions in Add Health, we found 5 in MLS. These included sexual abuse, physical abuse, parental separation/divorce or death, parental incarceration, and household substance abuse (father alcoholism). For all these questions, data were available from multiple time points and we assumed that ACE happened if an event was recorded in at least one time point. Further, the age at occurrence of the event was taken into account to ensure that the event happened before age 18. Moreover, parental separation/divorce or death and  parental incarceration was available from multiple sources (parents and children). We assumed that the event happened if at least one source showed it. To compute the final score for a given participant, we summed the responses for events with non-missing values and divided it by the number of events with non-missing response such that the final score is between 0-1.

In CHDS, we found the same 5 ACE-related questions as in MLS. So we followed a similar procedure as for MLS when data were available at multiple time points.

\section{Details on Model Re-calibration}

Calibration plot graphically represents agreement between the observed and predicted probabilities/counts \citep{janssen2008updating}. The calibration line in this plot is described by a calibration intercept and slope whose ideal values are 0 and 1, respectively. These parameters were estimated for the validation data sets by fitting a logistic regression model with the observed CUD status as the outcome variable and the score on the linear predictor ($\mathbf{X_{i}^\prime \mathbf{\beta}}$) obtained by applying the Add Health model as the sole predictor \citep{janssen2008updating}. 

We re-calibrated the Add Health model to adjust for differences in CUD prevalence between Add Health and the two validation data sets MLS, and CHDS \citep{moons2012risk}. We implement the simplest method of model re-calibration, updating only the intercept of the model. To this end, the calibration intercept was estimated by fitting a logistic regression model on a validation dataset with the intercept as the only free parameter and the linear predictor based on Add Health model as an offset term (that is the slope is fixed at 1). The calibration intercept was obtained using the CalibrationCurves R package \citep{calibrartion} which was then added to the original intercept of Add Health model. This updated the intercept of the Add Health model to adjust for the difference in the prevalence of CUD in validation data sets \citep{janssen2008updating,moons2012risk,van2019calibration,steyerberg2004validation}

\section{Robustness of Re-Calibration}
Since we used the validation data itself to estimate the calibration intercept, to check if the improvement in the E/O ratio is robust, we randomly divided each validation dataset into two subsets (training and test sets) through stratified sampling of cases and controls in a way such that each subset would have $50\%$ of cases and controls in that cohort. We used the training subset to estimate the calibration intercept and evaluated the performance of the resulting intercept-adjusted model on the test subset. Due to relatively small sample sizes, we repeated this process 20 times to account for sampling variability and obtained the mean calibration and discrimination statistics in the test subset for each validation cohort. The mean AUC obtained for the test subsets were 0.66 and 0.74 for MLS and CHDS, respectively. The observed E/O ratio after applying the intercept-adjusted model estimated to the test subsets were 0.99 for both the validation datasets. This suggests that the estimation of calibration intercept using validation data was relatively robust.

\bibliography{arxiv_supplementary}